\documentclass[aps,amsfonts,pra,twocolumn,showpacs]{revtex4-1}
\usepackage{epsfig,amsmath,amssymb,bm,epsf,graphicx,psfrag}
\usepackage[all]{xy}
\usepackage{color}

\def\bra#1{\langle#1\vert}
\def\ket#1{\vert#1\rangle}
\def\ketbra#1{\vert#1\rangle\langle#1\vert}

\def\ipr#1#2{\langle#1\vert#2\rangle}

\def\Longarrow{\protect\@lra}
\def\@lra{\relbar\joinrel\relbar\joinrel\relbar\joinrel%
          \relbar\joinrel\rightarrow}

\newcommand{\bc}{\begin{center}}
\newcommand{\ec}{\end{center}}
\newcommand{\be}{\begin{equation}}
\newcommand{\ee}{\end{equation}}
\newcommand{\bea}{\begin{eqnarray}}
\newcommand{\eea}{\end{eqnarray}}

\newcommand{\ncd}{\newcommand}
\ncd{\QCcns}{$QC_{\cal{C}}$}
\ncd{\QCc}{$QC_{\cal{C}}\;$}

\newtheorem{Thm}{Theorem}

\newtheorem{Lemma}{Lemma}

\definecolor{libl}{cmyk}{0.2,0.1,0,0}

\begin{document}

\title{The 2D AKLT state on the honeycomb lattice is a universal  resource for quantum computation}

\author{Tzu-Chieh Wei}
\affiliation{C. N. Yang Institute for Theoretical Physics, State
University of New York at Stony Brook, Stony Brook, NY 11794-3840,
USA}
\author{Ian Affleck}
\affiliation{Department of Physics and Astronomy, University of British
Columbia, Vancouver, British Columbia V6T 1Z1, Canada}
\author{Robert Raussendorf}
\affiliation{Department of Physics and Astronomy, University of British
Columbia, Vancouver, British Columbia V6T 1Z1, Canada}
\date{\today}

\begin{abstract}
Universal quantum computation can be achieved by simply performing
single-qubit measurements on a highly entangled resource state.
Resource states can arise from ground states of carefully designed
two-body interacting Hamiltonians. This opens up an appealing
possibility of creating them by cooling.  The family of
Affleck-Kennedy-Lieb-Tasaki (AKLT) states are the ground states of
particularly simple Hamiltonians with high symmetry, and their
potential use in quantum computation gives rise to a new research
direction. Expanding on our prior work [T.-C. Wei, I. Affleck, and
R. Raussendorf, Phys. Rev. Lett.~{\bf 106}, 070501 (2011)], we give
detailed analysis to explain why the spin-3/2 AKLT state on a
two-dimensional honeycomb lattice is a universal resource for
measurement-based quantum computation. Along the way, we also
provide an alternative proof that the 1D spin-1 AKLT state can be
used to simulate arbitrary one-qubit unitary gates. Moreover, we
connect the quantum computational universality of 2D random graph
states to their percolation property and show that these states
whose graphs are in the supercritical (i.e. percolated) phase are
also universal resources for measurement-based quantum computation.
\end{abstract}
\pacs{ 03.67.Ac,
03.67.Lx, 
64.60.ah,  
75.10.Jm 
} \maketitle

\section{Introduction}
The rules of quantum mechanics appear to perform certain tasks much
more efficiently than those of classical mechanics. The most
celebrated example is the factoring of a large integer by Shor's
quantum algorithm~\cite{Shor} that offers exponential speedup over
existing classical algorithms. Quantum computers that implement
generic quantum algorithms can take form in various computational
models, such as the standard circuit model~\cite{NielsenChuang00},
adiabatic quantum computer~\cite{AQC,AQC2}, and quantum
walk~\cite{Childs}, all of which proceed via the important feature
of quantum mechanics---the unitary evolution.

A different but equally powerful framework is the measurement-based
quantum
computation~\cite{GottesmanChuang,NielsenLeungChilds,Oneway}. A
particular computational model within this class is One-Way quantum
computation~\cite{Oneway} which we subsequently denote by MBQC. It
proceeds by single-qubit measurements alone on a highly entangled
initial resource state~\cite{Oneway,Oneway2,RaussendorfWei12}. For
MBQC, resource states that allow universal quantum computation turns
out to be very rare~\cite{Gross1}, but examples do
exist~\cite{Cluster, Gross, Verstraete,Cai}.  The first identified
universal resource state is the 2D cluster state on the square
lattice~\cite{Oneway,Cluster}. It was also shown that 2D cluster
states defined on regular lattices, such as triangular, hexagonal
and Kagom\'e, are also universal resources~\cite{Universal}. Cluster
states and related graph states can be created by the Ising
interaction from unentangled states~\cite{Cluster} and they have
been created with cold atoms in optical lattices~\cite{coldatom}.
However, they do not arise as unique ground states of two-body
interacting Hamiltonians~\cite{Nielsen}, although they can be an
approximate unique ground state~\cite{BartlettRudolph}. However, by
going beyond qubit systems and by careful design of Hamiltonians, a
few quantum states have been found that are both unique ground
states and universal for
MBQC~\cite{Chen,Cai10,WeiRaussendorfKwek11,LiEtAl}. This opens up an
alternative possibility of creating universal resource states by
cooling the systems.

Independently of the development on quantum computation, Affleck,
Kennedy, Lieb and Tasaki (AKLT) constructed a family of states that
were ground states of isotropic antiferromagnet-like
Hamiltonians~\cite{AKLT,AKLT2,AKLT3}. In any dimension, AKLT states
are ground states of  particularly simple Hamiltonians which only
have nearest-neighbor two-body interactions, are rotationally
invariant in spin space and shares all spatial symmetries of the
underlying lattice.  In particular, AKLT provided an explicit
example of a one-dimensional spin-1 chain that has a finite spectral
gap above the ground state, supporting Haldane's conjecture on
integer spin chains with spin rotation symmetry~\cite{Haldane}.
These valence-bond states turned out to be the first examples of
matrix product states (MPS)~\cite{MPS} and projected entangled pairs
states (PEPS)~\cite{Verstraete,PEPS}. The use of MPS and PEPS also
gives rise to a new perspective on MQBC~\cite{Verstraete,Gross}. In
particular, it was recently discovered that the one-dimensional
spin-1 AKLT state~\cite{AKLT,AKLT2} can serve as resources for
restricted computations~\cite{Gross,Brennen}, i.e., implementation
of arbitrary one-qubit rotations. The discovery of the
resourcefulness of AKLT states creates additional avenues for its
experimental realization~\cite{Resch}, and has instilled novel
concepts in MBQC, such as the renormalization group and the
holographic principle~\cite{Bartlett,Miyake}. However, to achieve
universal quantum computation within the measured-based architecture
a two-dimensional structure is needed.

In Ref.~\cite{Cai10}, Cai et al. considered stacking up 1D AKLT
quasichains to form a 2D structure. Their construction showed that
the resulting state, even though it is longer an AKLT state, can
provide universal quantum computation. Later independently by
us~\cite{WeiAffleckRaussendorf11} and by Miayke~\cite{Miyake10}, it
was shown that indeed the 2D AKLT state on the honeycomb lattice
provides a universal resource for MQBC. Here, expanding on our prior
work~\cite{WeiAffleckRaussendorf11}, we provide an alternative proof
that the 1D spin-1 AKLT state can be used to simulate arbitrary
one-qubit unitary gates, and  generalize the method and give
detailed analysis to the proof that the spin-3/2 AKLT state on a
two-dimensional honeycomb lattice is a universal resource for
measurement-based quantum computation. We do this by showing that a
2D cluster state can be distilled by local operations. Along the
way, we have connected the quantum computational universality of 2D
random graph states to their percolation property. We note that
extension of our approach using Positive Operator Valued Measure
(POVM) and percolation consideration to computational universality
have been successfully applied to a deformed AKLT model in
Ref.~\cite{DarmawanBrennenBartlett}.

The structure of the present paper is as follows. In
Sec.~\ref{sec:1D} we discuss how to locally convert a 1D AKLT to a
1D cluster state. In Sec.~\ref{sec:Reduction} we outline and
illustrate the method of how to locally convert the 2D AKLT state to
a random graph state. We then give the general proof in
Sec~\ref{sec:generalproof}. In Sec.~\ref{sec:RandomGraph} we show
the quantum computational universality of these graph states is
related to the percolation of the graph and show how to convert
these graph states to a 2D cluster state on a square lattice. We
support our assertion with Monte Carlo simulations in
Sec.~\ref{sec:MonteCarlo} and conclude in Sec.~\ref{sec:Conclude}.
In the Appendices, we use a different approach to obtain the
probability of getting any POVM outcomes.
\section{One dimension}
\label{sec:1D} We begin by investigating the 1D AKLT state and how
it can be locally converted to a 1D cluster state. By doing so, we
have thus proved the equivalence of the capability to simulate
one-qubit unitary gates for both types of states. Many of the
methods developed in this section can be extended to the more
interesting case of 2D AKLT state on the honeycomb lattice.
\subsection{1D spin-1 AKLT state and 1D cluster state}
\begin{figure}
   \includegraphics[width=6cm]{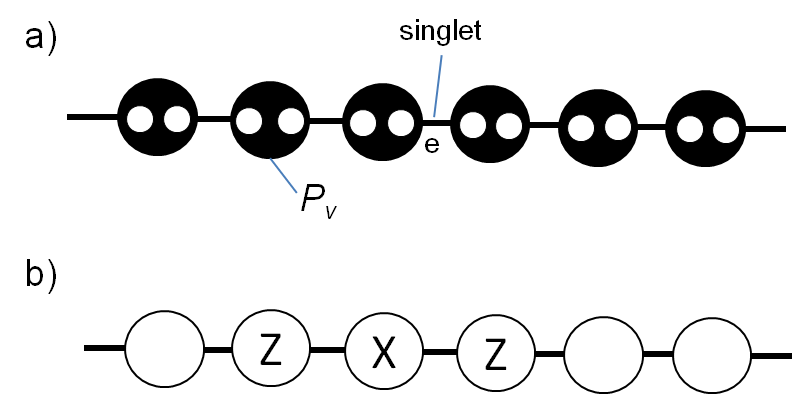}
  \caption{\label{fig:1D}The 1D AKLT state (a) and the 1D cluster state (b).}
\end{figure}
The 1D AKLT state~\cite{AKLT} can be understood by using the
valence-bond-solid (VBS) picture, as illustrated in
Fig.~\ref{fig:1D}a. (1) First one regards  a spin-1 particle at each
site as consisting of two virtual spin-1/2 particles (qubits), each
of which forms a singlet with the virtual qubit on the neighboring
site: $|\phi\rangle_e\equiv|01\rangle_e-|10\rangle_e$, where the
normalization is omitted,  $|0\rangle \equiv |\uparrow\rangle$ and
$|1\rangle \equiv |\downarrow\rangle$  are eigenstates of Pauli
$\sigma_z$, and $e$ denotes the edge that links the two virtual
qubits. (2) A local projection is then made at every site that maps
the state of the two virtual qubits to their symmetric subspace,
which is then identified as the Hilbert space of spin-1 particle:
\begin{eqnarray}
\!\!\!\!\!\!\!\!\!\!&&\hat{P}_{v}=\ket{S_z=1}\bra{00}+\ket{S_z=-1}\bra{11}+\ket{S_z=0}\bra{\psi^+}, \label{eqn:Pv1D}\\
\!\!\!\!\!\!\!\!\!\!&&\ket{\psi^+}\equiv\frac{1}{\sqrt{2}}(\ket{01}+\ket{10}),
\end{eqnarray}
where $\ket{S_z=\pm1,0}$ are the three $S=1$ angular momentum
eigenstates: $\hat{S}_z|S_z\rangle=S_z|S_z\rangle$. For convenience
we shall take the periodic boundary condition, so that the last site
of the 1D chain is actually connected to the first site of the
chain. Open boundary condition can be dealt with by attaching qubits
at the ends. The 1D AKLT is therefore given by
\begin{equation}
  \label{AKLT1D}
  |\Phi^{(1D)}_{\rm AKLT}\rangle\equiv\bigotimes_{v }\hat{P}_{v}
\bigotimes_{e} |\phi\rangle_e,
\end{equation}
which is the unique ground state of the following spin-isotropic
Hamiltonian with a finite gap~\cite{AKLT,AKLT2}:
\begin{equation}
H_{\rm AKLT}^{S=1}=\sum_v\Big[ \vec{S}_v\cdot
\vec{S}_{v+1}+\frac{1}{3}(\vec{S}_v\cdot
\vec{S}_{v+1})^2+\frac{2}{3} \Big],
\end{equation}
where $\vec{S}_v$ denotes the vector of the spin operator at site
$v$. This AKLT model provided strong evidence in support of the
Haldane's conjecture~\cite{Haldane}.

On the other hand, the 1D cluster state $|{\cal C}_{1D}\rangle$ also
can be understood similarly by projecting virtual entangled pairs to
physical spins, known as projected entangled pairs states
(PEPS)~\cite{PEPS}, where the virtual entangled pair is replaced by
$|\phi_H\rangle_e\equiv |00\rangle+|01\rangle+|10\rangle-|11\rangle$
and the local projection is given by $\hat{P}_v^{{\cal
C}}\equiv|0\rangle\langle 00|+|1\rangle\langle11|$, giving rise to
\begin{equation}
|{\cal C}^{(1D)}\rangle\equiv\bigotimes_{v }\hat{P}_{v}^{{\cal C}}
\bigotimes_{e} |\phi_H\rangle_e.
\end{equation}
 However, for our
purpose, it will be useful to define equivalently the cluster state
as the common eigenstate of the following operators:
\begin{equation}
\label{eqn:1Dcluster}
 Z_{v-1} X_v Z_{v+1}|{\cal
C}^{(1D)}\rangle=|{\cal C}^{(1D)}\rangle,
\end{equation}
for all sites $v$, where $v\pm 1$ are the two neighboring sites of
$v$ on the chain. Note that for convenience we denote the three
Pauli matrices by $X\equiv \sigma_x$, $Y\equiv\sigma_y$ and
$Z\equiv\sigma_z$, and use the two notations interchangeably.
Moreover, the choice of ``+1'' or ``-1'' eigenvalue is arbitrary, as
the resulting states are related by local unitary transformation.
The 1D cluster state can be used to simulate one-qubit unitary
operation on one qubit and is the basic ingredient in
MBQC~\cite{Oneway}.

In fact, the 1D AKLT state has been shown to be able to simulate
one-qubit unitary operation as the 1D cluster
state~\cite{Gross,Brennen,Miyake} by explicitly constructing
one-qubit universal gates. It has also been realized that the spin-1
AKLT state can actually be converted, via local operations, to the
1D spin-1/2 cluster state with a random length~\cite{Chen10}. In the
following section, we provide an alternative method for the
reduction of the 1D AKLT state to a 1D cluster state. This method
will then be generalized later for the reduction of the 2D AKLT
state.

\subsection{Reducing 1D AKLT state to a 1D cluster state}
As spin-1 Hilbert space is of
dimensionality three, in order to convert to dimensionality two of a
qubit, a projection or a generalized measurement is needed. In the
mapping $\hat{P}_v$ in Eq.~(\ref{eqn:Pv1D}), there is a
two-dimensional subspace spanned by $|S=1,S_z=1\rangle$ and
$|S=1,S_z=-1\rangle$ or equivalently by the two virtual qubits
$|00\rangle$ and $|11\rangle$. One can therefore consider
\begin{equation}
F_z= (\ketbra{S_z=1}+\ketbra{S_z=-1})/\sqrt{2}
\end{equation}
 as a
projection that preserves a two-dimensional subspace, where we
suppress the label $S=1$. However, what happens if the projection is
not successful and it ends up in the subspace orthogonal to that
spanned by $\ket{S_z=1}$ and $\ket{S_z=-1}$? To solve this
``leakage'' problem, one takes  advantage of the rotation
symmetry and adds two more projections:
\begin{eqnarray}
\!\!\!\!\!\!\!\!\!\!&&F_x = (\ketbra{S_x=1}+\ketbra{S_x=-1})/\sqrt{2}, \\
\!\!\!\!\!\!\!\!\!\!&&F_y=
(\ketbra{S_y=1}+\ketbra{S_y=-1})/\sqrt{2},
\end{eqnarray}
and
notice  the completeness relation in the spin-1 Hilbert space:
\begin{equation} \sum_{\alpha=x,y,z}
F^\dagger_\alpha F_\alpha =\openone_{S=1}.
\end{equation}

The above $F$'s constitute the so-called generalized measurement or
POVM, characterized by $\{F_\alpha^\dagger F_\alpha\}$. Their
physical meaning is to define a two-dimensional subspace and to
specify a preferred quantization axis $x$, $y$ or $z$. In principle,
the POVM can be realized by a unitary transformation $U$ jointly on
a spin-1 state, denoted by $|\psi\rangle$, and a meter state
$|0\rangle_m$ such that
\begin{equation}
U |\psi\rangle|0\rangle_m = \sum_{\alpha}
F_\alpha|\psi\rangle|\alpha\rangle_m,
\end{equation}
where for the meter states $\langle
\alpha|\alpha'\rangle=\delta_{\alpha,\alpha'}$. A measurement on the
meter state will result in a random outcome $\alpha$, for which the
spin state is projected to
$F_\alpha|\psi\rangle$~\cite{NielsenChuang00}.

{\bf Claim.} We shall show that after performing the generalized
measurement on all sites with $\{a_v\}$ denoting the measurement
outcome the resulting state
\begin{equation}
\label{eqn:1Dpost} |\psi(\{a_v\})\rangle\equiv \bigotimes_{v}
F_{v,a_v} |\Phi^{(1D)}_{\rm AKLT}\rangle
\end{equation}
is an ``encoded'' 1D cluster state.

In the following we shall make use of the equivalent representation
of the AKLT state by the virtual qubits; see Eq.~(\ref{eqn:Pv1D}),
e.g., $|S_z=1\rangle =|00\rangle$ and $|S_z=-1\rangle=|11\rangle$,
where the r.h.s are two-qubit states. In this regard, we can think
of $F$ operators in terms of two-qubit operators:
\begin{subequations}
\label{eqn:F1D}
\begin{eqnarray}
\tilde{F}_z&=&(\ketbra{00}+\ketbra{11})/\sqrt{2},\\
\tilde{F}_x&=&(\ketbra{++}+\ketbra{--})/\sqrt{2},\\
\tilde{F}_y&=&(\ketbra{i,i}+\ketbra{-i,-i})/\sqrt{2},
\end{eqnarray}
\end{subequations}
where $|\pm\rangle$ satisfy $\sigma_x|\pm\rangle=\pm|\pm\rangle$ and
$|\pm i\rangle$ satisfy $\sigma_y|\pm i\rangle=\pm|\pm i\rangle$.
Thus, in terms of these $\tilde{F}$'s the post-measurement
state~(\ref{eqn:1Dpost}) is simply given by
\begin{equation}
\label{eqn:1Dpost2} |\psi(\{a_v\})\rangle\equiv \bigotimes_{v}
F_{v,a_v} \bigotimes_e |\phi\rangle_e.
\end{equation}
 Naturally as with $|S_z=\pm1\rangle$, there is also the correspondence
between the other two $S=1$ states and the two-qubit states in $x$
and $y$ bases: $|S_x=1\rangle=|++\rangle$,
$|S_x=-1\rangle=|--\rangle$, $|S_y=1\rangle=|i,i\rangle$, and
$|S_y=-1\rangle=|-i,-i\rangle$. The use of qubit enables us to take
the advantage of the stabilizer formalism~\cite{Stabilizer}, even
though its use is not essential.

\begin{figure}
   \includegraphics[width=8cm]{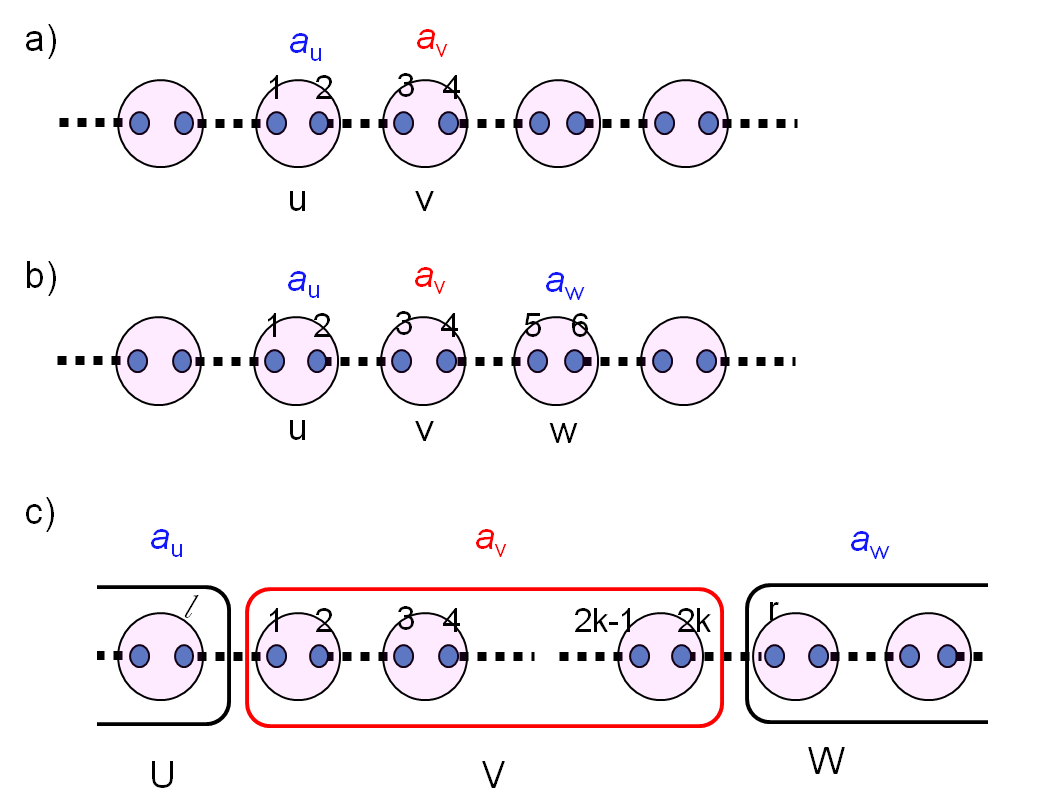}
  \caption{\label{fig:figZXZ} Illustration of (a) encoding and (b)\& (c) stabilizer operator.
}
\end{figure}

First, let us explain the meaning of ``encoding'' used in the claim.
Suppose two neighboring sites $u$ and $v$ have the same outcome
$a=z$; see Fig.~\ref{fig:figZXZ}a. Then the two-dimensional
subspaces at sites $u$ and $v$ are both spanned by $|00\rangle$ and
$|11\rangle$. However, the singlet state between the two virtual
qubits connecting $u$ and $v$ dictates that the appearance of the
basis states are anti-correlated. For example, $|(00)_u\rangle$ at
site $u$ cannot coexist with $|(00)_v\rangle$ at site $v$. There are
only two possible basis states for the two sites $u$ and $v$:
$|``0"\rangle\equiv|(00)_u(11)_v\rangle$ and
$|``1"\rangle\equiv|(11)_u(00)_v\rangle$. These two states
``encode'' a logical qubit $\{|``0"\rangle, |``1"\rangle\}$. In
terms of spin-1 notation, they are $|S_z=1, S_z=-1\rangle_{u,v}$ and
$|S_z=-1, S_z=1\rangle_{uv}$, showing the antiferromagnetic
properties of the AKLT state. From these two states, logical Z and X
operators can be defined: $Z\equiv \ketbra{``0"}-\ketbra{``1"}$ and
$X\equiv |``0"\rangle\langle ``1"| + |``1"\rangle\langle ``0"|$ (and
thus the Pauli operator $Y=-i ZX$ can be determined). In the same
manner, for $k$ consecutive sites with the same outcome $a$, only
one qubit is encoded by the $k$ physical spins with quantization
axis being in the $a$-direction. We shall refer to these sites
collectively as a {\it domain\/}. On the other hand, for two
neighboring sites having different outcome $a_u\ne a_v$, the four
combination $|S_{a_u}=\pm1, S_{a_v}=\pm1\rangle$ can appear and each
site is effectively a qubit.

The above analysis can be expressed in terms of the stabilizer
formalism. In the example that neighboring $u$ and $v$ share the
same outcome $a=z$ (i.e., the domain consists of two sites $u$ and
$v$), for the two virtual qubits of site $u$ (denoted by the $1$ and
$2$) we have $\sigma_z^{[1]}\otimes \sigma_z^{[2]}
\tilde{F}_{u,z}=\tilde{F}_{u,z}$. This means $\sigma_z^{[1]}\otimes
\sigma_z^{[2]} |\psi(\{a_v\})\rangle= |\psi(\{a_v\})\rangle$.
Similarly, for site $v$ (with two virtual qubits labeled as $3$ and
$4$) we have $\sigma_z^{[3]}\otimes \sigma_z^{[3]}
|\psi(\{a_v\})\rangle= |\psi(\{a_v\})\rangle$. However, because of
the singlet between $2$ and $3$, we have $-\sigma_z^{[2]}\otimes
\sigma_z^{[3]} |\psi(\{a_v\})\rangle= |\psi(\{a_v\})\rangle$. The
above three operators $\{\sigma_z^{[1]}\otimes
\sigma_z^{[2]},\sigma_z^{[3]}\otimes
\sigma_z^{[4]},-\sigma_z^{[2]}\otimes \sigma_z^{[3]}\}$ are called
the stabilizer generators and they define the logical qubit basis
states: $|``0"\rangle\equiv|(00)_u(11)_v\rangle$ and
$|``1"\rangle\equiv|(11)_u(00)_v\rangle$, as one can verify that
they are the common eigenstates of these operators with eigenvalue
$+1$. The stabilizer generators are effectively identity operators
in the logical-qubit Hilbert space. To define the logical $Z$
operator, there are many equivalent choices: e.g., $\sigma_z^{[1]}$,
$\sigma_z^{[2]}$, $-\sigma_z^{[3]}$ and  $-\sigma_z^{[4]}$. Any of
them can be taken to one another by multiplication of some
combination of the stabilizer generators. To complete the logical
qubit operators, the $X$ operator can be taken as $X\equiv
\sigma_x^{[1]}\sigma_x^{[2]}\sigma_x^{[3]}\sigma_x^{[4]}$, which
flips $|``0"\rangle$ to $|``1"\rangle$, and vice versa. Other
outcomes can be dealt with in a similar way, and these are
summarized in Table~\ref{tbl:coding1D}. Two important properties are
that (i) each domain can contain more than one physical qubit and is
only one logical qubit; (2) the qubit basis depends on the shared
outcome of the POVM.

\begin{table}
  \begin{tabular}{l|r|r|r}
    \parbox{1.4cm}{POVM outcome\vspace{1mm}} & \multicolumn{1}{c|}{$z$} & \multicolumn{1}{c|}{$x$} & \multicolumn{1}{c}{$y$} \\ \hline
    \parbox{1.6cm}{stabilizer generator} & $\lambda_{i}\lambda_{i+1} \sigma_z^{[i]}\sigma_z^{[i+1]}$ & $\lambda_{i}\lambda_{i+1} \sigma_x^{[i]}\sigma_x^{[i+1]}$ & $\lambda_{i}\lambda_{i+1}\sigma_y^{[i]}\sigma_y^{[i+1]}$ \\
    $\overline{X}$ & $\bigotimes_{j = 1}^{2|{\cal{C}}|} \sigma_x^{[j]}$ & $\bigotimes_{j = 1}^{2|{\cal{C}}|} \sigma_z^{[j]}$ & $\bigotimes_{j = 1}^{2|{\cal{C}}|} \sigma_z^{[j]}$ \\
    $\overline{Z}$ & $\lambda_i \sigma_z^{[i]}$ & $\lambda_i \sigma_x^{[i]}$ & $\lambda_i \sigma_y^{[i]}$
  \end{tabular}
 \caption{\label{tbl:coding1D} The dependence of stabilizers and encodings for the
 random graphs
  on the local POVM outcome in the case of 1D AKLT state. $|{\cal C}|$ denotes the total number of virtual qubits contained in a
  domain. In the first line, $i = 1\, ..\, 2|{\cal{C}}|-1$, and in the third line
  $i = 1\,..\, 2|{\cal{C}}|$.
  One choice of the sign is $\lambda_i = 1$ if the virtual qubit $i$ is on the odd number of sites (relative to,  e.g., the left end)
  in the domain  and $\lambda_i = -1$ otherwise.}
\end{table}

We remark that even though a domain may contain two or more sites
one can perform projective measurement on all but one site in the
basis defined by $\{\ket{S_a=1}\pm \ket{S_a=-1}\}$, where $a$ is the
label of the POVM outcome for the domain. The domain is then reduced
to a single site but still preserves the same degree of entanglement
with its neighbors.

To show that the post-POVM state is an (encoded) cluster state, let
us illustrate with the example shown in Fig.~\ref{fig:figZXZ}b. Let
us label the three sites by $u$, $v$ and $w$ respectively. Suppose
the POVM outcomes on these sites are $a_u=x$, $a_v=z$, and $a_w=x$,
respectively. First note that $-\sigma_x^{[2]}\sigma_x^{[3]}$
commutes with $\tilde{F}_{u,x}$ and $-\sigma_x^{[4]}\sigma_x^{[5]}$
commutes with $\tilde{F}_{w,x}$. Note also that
$-\sigma_x^{[2]}\sigma_x^{[3]}$ is a stabilizer operator of the
singlet between $2$ and $3$, but it does not commute with
$\tilde{F}_{v,z}$. Similarly, $-\sigma_x^{[4]}\sigma_x^{[5]}$ is a
stabilizer operator of the singlet between $4$ and $5$, but it does
not commute with $\tilde{F}_{v,z}$, either.  However, if we multiply
all the above operators, we obtain
\begin{equation} K_v\equiv \sigma_x^{[2]}
\sigma_x^{[3]}\sigma_x^{[4]}\sigma_x^{[5]}.
\end{equation}
Because $\sigma_x^{[3]}\sigma_x^{[4]}$ commutes with
$\tilde{F}_{v,z}$, due to the identity
\begin{eqnarray}
&&\sigma_x\otimes\sigma_x
(|00\rangle\langle00|+|11\rangle\langle11|)\nonumber \\
& =&(|11\rangle\langle
00|+|00\rangle\langle11|)\nonumber\\
&=& (|00\rangle\langle00|+|11\rangle\langle11|)\,
\sigma_x\otimes\sigma_x,
\end{eqnarray}
$K_v$ is thus a stabilizer operator for the post-POVM state. In
terms of logical Pauli operators $\overline{Z}_u\equiv
\sigma_x^{[2]}$ , $\overline{Z}_w\equiv \sigma_x^{[5]}$,
$\overline{X}_v\equiv \sigma_x^{[3]}\sigma_x^{[4]}$, we arrive at
the stabilizer operator $K_V=\overline{Z}_U \overline{X}_V
\overline{Z}_W$. This is the stabilizer operator defining a linear
cluster state; see Eq.~(\ref{eqn:1Dcluster}).

 As a further illustration, let
us consider  the same three sites in Fig.~\ref{fig:figZXZ}b but with
$a_u=x$, $a_v=z$, and $a_w=y$, i.e., the last site has a different
outcome $a_w=y$ than the above example. Because of this, one now
considers $-\sigma_y^{[4]}\sigma_y^{[5]}$ instead of
$-\sigma_x^{[4]}\sigma_x^{[5]}$ and can show that the following
operator is a stabilizer generator:
\begin{equation}
 K_V\equiv \sigma_x^{[2]}
\sigma_x^{[3]}\sigma_y^{[4]}\sigma_y^{[5]}.
\end{equation}
Now, we use the logical operators ${Z}_u\equiv \sigma_x^{[2]}$,
$\overline{Z}_w\equiv \sigma_y^{[5]}$,  $\overline{X}_v\equiv
\sigma_x^{[3]}\sigma_x^{[4]}$, and $\overline{Z}_v\equiv
\sigma_z^{[4]}$ and we arrive at
\begin{equation}
K_V=\overline{Z}_u (i\overline{X}_v \overline{Z}_v)\overline
{Z}_w=\overline{Z}_u \overline{Y}_v \overline{Z}_w.\end{equation}
 Although the stabilizer operator $K_v$ is not
of the canonical form of the cluster-state stabilizer
$\overline{Z}_u \overline{X}_v \overline{Z}_w$, they are related by
local unitary transformation that leaves $\overline{Z}_v$ invariant.

\subsection{General proof of 1D encoded cluster state}
 The examples in the previous section prepare us
for the general proof that the post-POVM state is an encoded 1D
cluster state. Consider Fig.~\ref{fig:figZXZ}c, in which there are
three blocks labeled by $U$, $V$, and $W$, that may contain multiple
sites having same POVM outcome, $a_u$, $a_v$, and $a_w$,
respectively. Let us label the last virtual qubit in block $U$ by
$l$, the first virtual qubit in block $W$ by $r$, and the virtual
qubits in block $V$ by $1,2,\ldots,2k$. Because $a_v\ne a_u$ and
$a_v\ne a_w$, we can separate the proof into two cases: (1)
$a_u=a_w$, just as the first example $(a_u,a_v,a_w)=(x,z,x)$ given
in last section; (2) $a_u\ne a_w$, just as the second example
$(a_u,a_v,a_w)=(x,z,y)$ given in last section. The proof given below
is a straightforward generalization of these examples.

Case (1). Let us define $a\equiv a_w=a_u$. For the edges connecting
$V$ to $U$ and to $W$, consider the two operators:
$-\sigma_{a}^{[l]}\sigma_{a}^{[1]}$ and
$-\sigma_{a}^{[2k]}\sigma_{a}^{[r]}$. Denote by $\alpha$ the label
such that $\overline{X}_V\equiv\otimes_{j=1}^{2k}
\sigma_\alpha^{[j]}$ is the logical X operator for the block $V$.
For the edges connecting virtual qubits inside $V$, consider the
operator:
$\sigma_\alpha^{[2]}\sigma_\alpha^{[3]}..\sigma_\alpha^{[2k-1]}$.
The product of these three operators can be verified to be the
stabilizer operator for the post-POVM state:
\begin{equation}
K_V\equiv
\sigma_{a}^{[l]}\sigma_{a}^{[1]}\sigma_{\alpha}^{[1]}\overline{X}_V\sigma_{\alpha}^{[2k]}\sigma_{a}^{[2k]}\sigma_{a}^{[r]}.
\end{equation}
As $a\ne a_v$, either $\sigma_a=\sigma_\alpha$ or $\sigma_a=\pm i
\sigma_\alpha \sigma_{a_v}$ and thus either
$\sigma_a\sigma_\alpha=\openone$ or $\mp i\sigma_{a_v}$, with $\pm \sigma_{a_v}$ being  a logical $Z$ for block $V$ (and
$\overline{Z}^2=\openone$ from contributions of virtual qubits 1 and $2r$ ). Using the encoded $\overline{Z}$ for
block $U$ and $W$, i.e., $\overline{Z}_U=\pm \sigma_{a}^{[l]}$ and
$\overline{Z}_W=\pm\sigma_{a}^{[r]}$, we have
\begin{equation}
K_V= \pm \overline{Z}_U \overline{X}_V \overline{Z}_W
\end{equation}
is a stabilizer operator. (The choice of $\pm$ depends on the
convention; see Table~\ref{tbl:coding1D}.)

Case (2).  For the edges connecting $V$ to $U$ and to $W$, consider
the two operators: $-\sigma_{a_u}^{[l]}\sigma_{a_u}^{[1]}$ and
$-\sigma_{a_w}^{[2k]}\sigma_{a_w}^{[r]}$. Denote by $\alpha$ the
label such that $\overline{X}_V=\otimes_{j=1}^{2k}
\sigma_\alpha^{[j]}$ is the logical X operator for the block $V$.
For the edges connecting virtual qubits inside $V$, consider the
operator:
$\sigma_\alpha^{[2]}\sigma_\alpha^{[3]}..\sigma_\alpha^{[2k-1]}$.
The product of these three operators is the stabilizer operator for
the post-POVM state:
\begin{equation}
K_V\equiv
\sigma_{a_u}^{[l]}\sigma_{a_u}^{[1]}\sigma_{\alpha}^{[1]}\overline{X}_V\sigma_{\alpha}^{[2k]}\sigma_{a_w}^{[2k]}\sigma_{a_w}^{[r]}.
\end{equation}
As $a_u\ne a_w$, $\alpha$ is either equal to $a_u$ or $a_w$ and
hence the product of
$\sigma_{a_u}^{[1]}\sigma_{\alpha}^{[1]}\sigma_{\alpha}^{[2k]}
\sigma_{a_w}^{[2k]}$ becomes either
$\sigma_{a_u}^{[1]}\sigma_{\alpha}^{[1]}$ or $\sigma_{\alpha}^{[2k]}
\sigma_{a_w}^{[2k]}$. Either of them is a logical $\pm i
\overline{Z}_V$. Thus, we have
\begin{equation}
K_V= \pm i \overline{Z}_U (\overline{Z}_V \overline{X}_V)
\overline{Z}_W=\pm \overline{Z}_U \overline{Y}_V \overline{Z}_W
\end{equation}
is a stabilizer operator. This concludes the proof that the
post-POVM state is an encoded 1D cluster state.

\subsection{Probability of a POVM outcome}
\label{sec:1Dprob} Given a set of POVM outcome $\{a_v\}$, what is
the probability $p(\{a_v\})$ that this occurs? This is can be
obtained from the norm square of the resulting {\it un-normalized\/}
post-POVM state $|\psi(\{a_v\})\rangle$, namely,
\begin{equation}
p(\{a_v\})=\langle\psi(\{a_v\})|\psi(\{a_v\})\rangle/\langle\Phi^{(1D)}_{\rm
AKLT}|\Phi^{(1D)}_{\rm AKLT}\rangle.
\end{equation}
Let us denote by $|V|$ the total number of domains, which is the
number of logical qubits and $|{\cal E}|$ the total number of edges
connecting domains. As we consider the periodic boundary condition,
trivially $|{\cal E}|=|V|$, except when all sites have the same POVM
outcome, i.e., $a_v=a$ for all $v$. Note that this latter case can
never occur if the total number of the original spins is odd, as the
frustrated configurations $|+1,-1,+1,-1,...,+1\rangle$ and
$|-1,+1,-1,+1,...,-1\rangle$ (with the first and last sites being
connected next to each other) cannot appear~\cite{AKLT,AKLT2}. It
turns out that, barring the exception of zero probability,
$p(\{a_v\})\sim 2^{|V|-|{\cal E}|}$. This is because for contracting
$\langle\psi(\{a_v\})|\psi(\{a_v\})\rangle$ to compute the norm we
need to evaluate $|\langle\alpha,\beta|(|01\rangle-|10\rangle)|^2$,
where $\alpha, \beta$ can be any of the six possibilities:
$\{0,1,+,-, +i, -i\}$. The ratio of the above expressions in the case were
$\alpha$ and $\beta$ belong to different bases to the case
where they belong to  the same
basis (thus $\alpha=-\beta$) is $1/2$. In total, there are $2^{|V|}$
terms of equal contribution to the norm square, each reduced by a
factor $2^{-|{\cal E}|}$. This results in the probability
$p(\{a_v\})\sim 2^{|V|-|{\cal E}|}$.

For the total number of sites $n$ being even, all the $3^n$ possible
POVM outcomes can occur, each with probability $p_0$, except for the
three configurations (with all $a_v$ being the same) having
probability $2p_0$. Solving $(3^n-3)p_0 + 3\cdot 2p_0=1$, we obtain
$p_0= 1/(3^n +3)$. For $n$ being odd, the three configurations with
all $a_v$ being the same cannot occur. All other configurations
occur with a probability $1/(3^n-3)$ each. For large $n$, it is a
very good approximation to regard all configurations $\{a_v\}$ as
occurring with equal probability and hence the resulting 1D cluster
state contains on average $2n/3$ qubits, which agrees with the
result in Ref.~\cite{Chen10}.
\section{Reduction of the 2D AKLT state}
\label{sec:Reduction}

Now that we have understood the 1D case, to show that the 2D AKLT
state is a universal resource for quantum computation, we proceed in
three steps. First, we show that it can be mapped to a random planar
graph state $|G({\cal{A}})\rangle$ by local generalized measurement,
with the graph $G({\cal{A}})$ depending on the set ${\cal A}$ of
measurement outcomes on all sites. Second, we show that the
computational universality of a typical resulting graph state
$|G({\cal{A}})\rangle$ hinges solely on the connectivity of
$G({\cal{A}})$, and is thus a percolation problem. Third, we
demonstrate through Monte Carlo simulation that the typical graphs
$G({\cal{A}})$ are indeed deep in the connected phase. We remark
that extension of our approach using POVM and percolation
consideration have been applied to a deformed AKLT model in
Ref.~\cite{DarmawanBrennenBartlett}.

\begin{figure}
 \includegraphics[width=8cm]{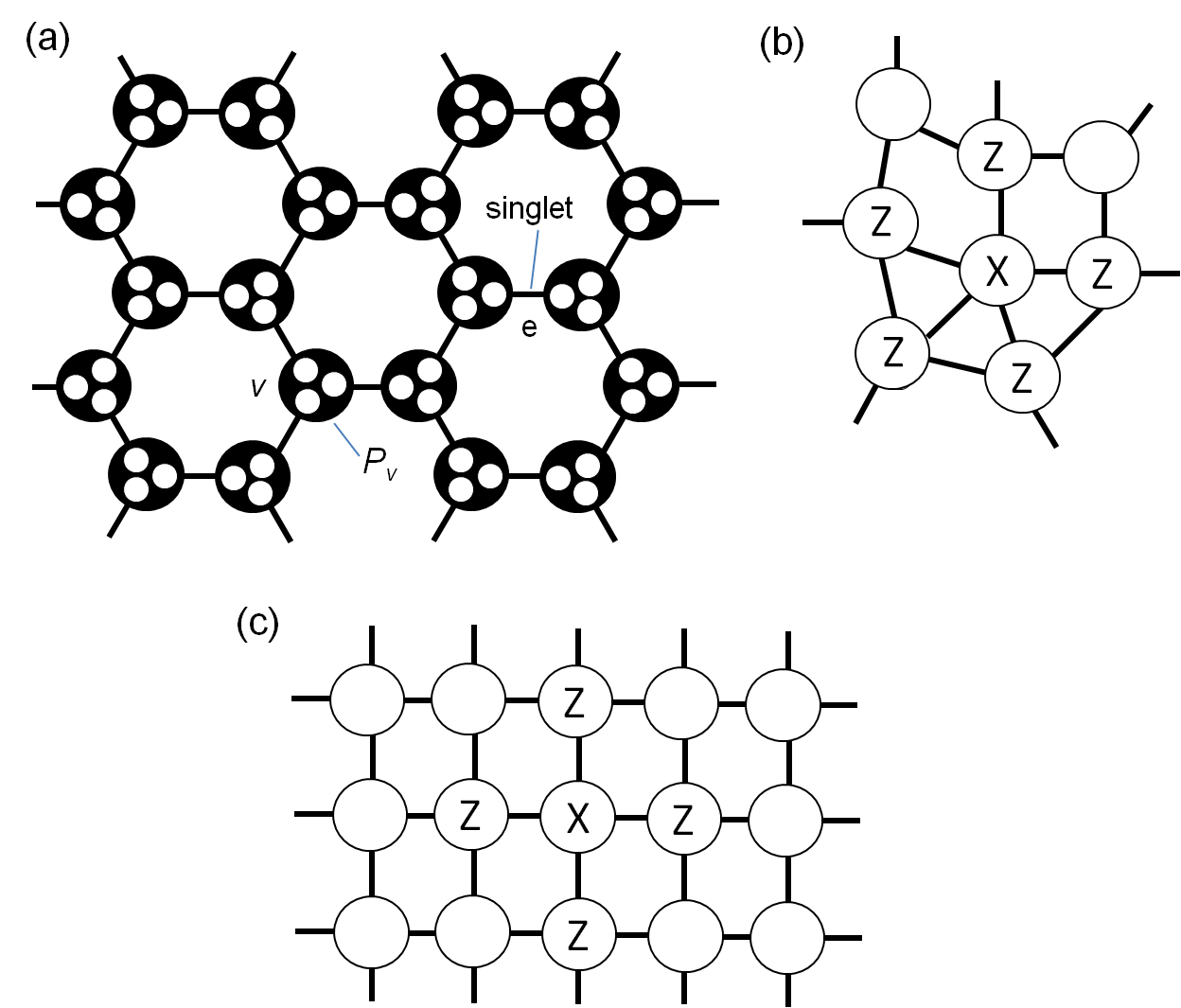}
  \caption{\label{fig:states}
  Illustrations of the AKLT state on the honeycomb lattice, a graph state and the 2D cluster state on
  a square lattice. (a) AKLT state. Spin singlets of two virtual spins 1/2 are located
  on the edges of the honeycomb lattice. A projection $P_{S,v}$ at each lattice site $v$
  onto the symmetric subspace of three virtual spins creates the AKLT state.
  (b) A graph state. One qubit (i.e. spin 1/2) resides at vertex of the graph. One stabilizer generator of the form $X_v\bigotimes_{u\in {\rm nb}(v)}
Z_u$ is shown. (c) 2D Cluster state is a special case of graph
states, where the graph is a two-dimensional square lattice.
  }
\end{figure}

The AKLT state~\cite{AKLT,AKLT2} on the honeycomb lattice ${\cal{L}}$ has one
spin-3/2 per site of ${\cal{L}}$. The state space of each spin 3/2 can be
viewed as the symmetric subspace of three virtual spin-1/2's, i.e., qubits. In
terms of these virtual qubits, the AKLT state on ${\cal{L}}$ is
\begin{equation}
  \label{AKLT2}
  |\Phi_{\rm AKLT}\rangle\equiv\bigotimes_{v \in V({\cal{L}})}P_{S,v}
\bigotimes_{e \in E({\cal{L}})} |\phi\rangle_e,
\end{equation}
where $V({\cal L})$ and $E({\cal L})$  denote the set of vertices
and edges of ${\cal L}$, respectively. $P_{S,v}$ is the projection
onto the symmetric (equivalently, spin 3/2) subspace at site $v$ of
${\cal{L}}$:
\begin{equation}
P_S\equiv
\ketbra{000}+\ketbra{111}+\ketbra{W}+\ketbra{\overline{W}},
\end{equation}
where
\begin{eqnarray}
\ket{W}&\equiv&\frac{1}{\sqrt{3}}(\ket{001}+\ket{010}+\ket{100}),\\
\ket{\overline{W}}&\equiv&\frac{1}{\sqrt{3}}(\ket{110}+\ket{101}+\ket{011}).
\end{eqnarray}
The mapping between three virtual qubits and spin-3/2 is given by:
$\ket{000}\leftrightarrow \ket{3/2,3/2}$, $\ket{111}\leftrightarrow
\ket{3/2,-3/2}$, $\ket{W}\leftrightarrow\ket{3/2,1/2}$ and
$\ket{\overline{W}}\leftrightarrow\ket{3/2,-1/2}$.
 For an edge $e=(v,w)$, $|\phi\rangle_{e}$ denotes a
singlet state, with one spin 1/2 at vertex $v$ and the other at $w$.
For illustration, see Fig.~\ref{fig:states}a. The AKLT state is the
ground state of the following Hamiltonian:
\begin{equation}
 H_{AKLT}^{S=3/2}=\!\!\sum_{{\rm edge}\,\langle i,j\rangle}\Big[ \vec{S}_i\cdot \vec{S}_{j}+\frac{116}{243}(\vec{S}_i\cdot \vec{S}_{j})^2+\frac{16}{243}(\vec{S}_i\cdot
 \vec{S}_{j})^3\Big],
\end{equation}
where an irrelevant constant term has been dropped.

Next, we give the definition of a graph state~\cite{Hein}, to which we shall
prove that the AKLT state can be locally converted. A graph state $\ket{G}$ is
a stabilizer state~\cite{Stabilizer} with one qubit per vertex of the graph
$G$. It is the unique eigenstate of a set of commuting
operators~\cite{Cluster}, usually called the stabilizer generators,
\begin{equation}
\label{eqn:stabilizerGen} X_v\bigotimes_{u\in {\rm nb}(v)}
Z_u\,\, \ket{{G}}=\ket{{G}}, \ \forall v\in V({G}),
\end{equation}
where ${\rm nb}(v)$ denotes the neighbors of vertex $v$, and $X\equiv
\sigma_x$, $Y\equiv\sigma_y$ and $Z\equiv\sigma_z$ are the three Pauli
matrices. A cluster state is a special case of graph states, with the
underlying graph being a regular lattice; see e.g. Fig.~\ref{fig:states}b for
the illustration. Any 2D cluster state is a universal resource for
measurement-based quantum computation~\cite{Oneway,Universal}.

\begin{figure}
  \hspace*{0.1cm} \includegraphics[width=8cm]{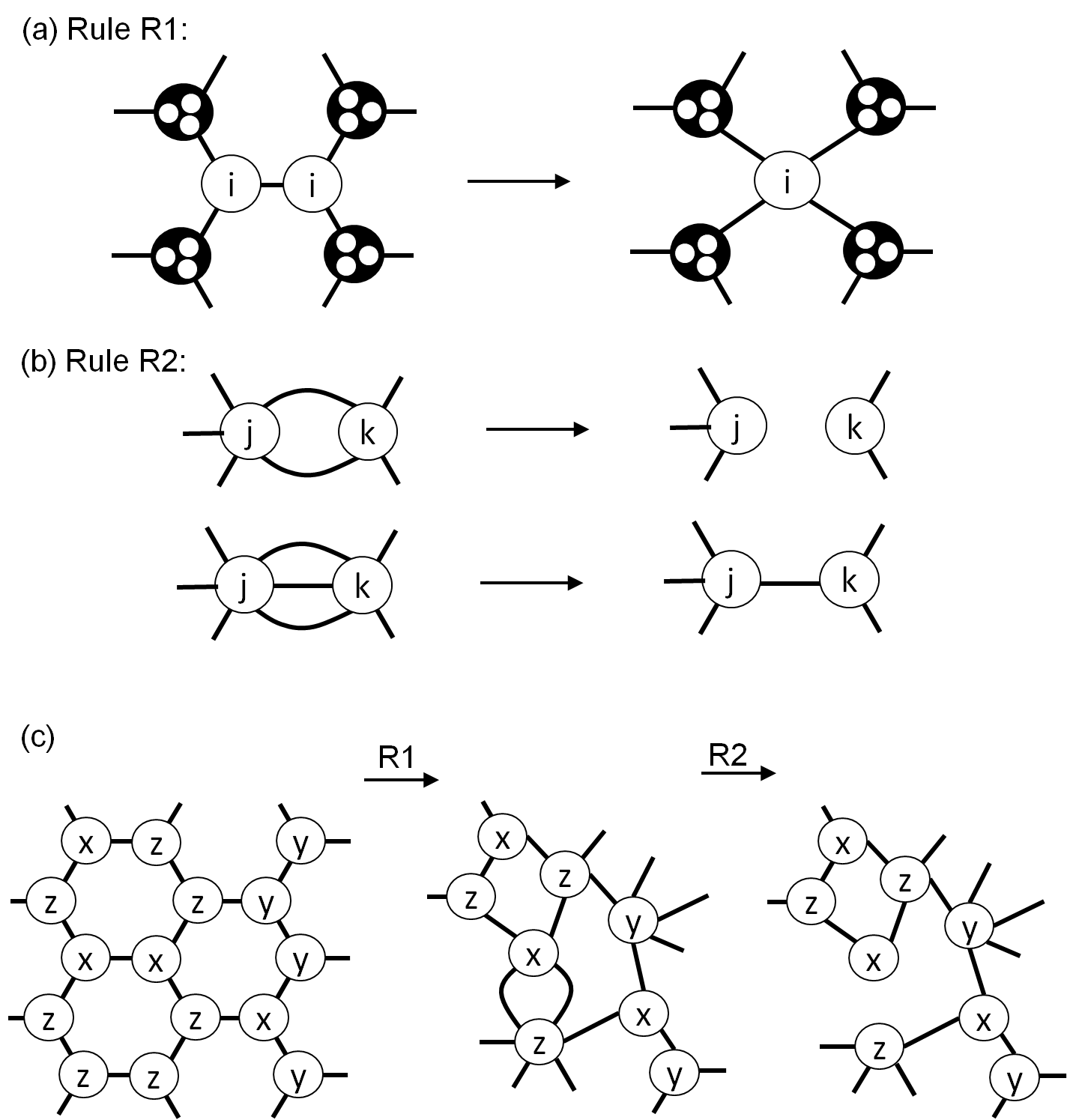}
  \caption{\label{merge}Graphical rules for transformation of the lattice ${\cal{L}}$
  into the graph $G({\cal{A}})$, depending on the POVM outcomes ${\cal{A}}$.
  (a) Graph Rule 1: A single edge between neighboring sites with the same POVM outcome
  $i \in \{x,y,z\}$ is contracted. (b) Rule 2: Pairs of edges between the same vertices are deleted.
  (b) An example for the application of the rules 1 and 2 to a small honeycomb
  lattice. The alphabets ($j\ne k$) inside the circle indicate the POVM outcomes.
  (c) An example to illustrate the applications of the graph rules. }
\end{figure}

To show that the 2D AKLT state of four-level spin-3/2 particles can
be converted to a graph state of two-level qubits, we need to
preserve a local two-dimensional structure at each site. This is
achieved by a local generalized measurement~\cite{NielsenChuang00},
also called positive-operator-value measure (POVM), on every site
$v$ on the honeycomb lattice ${\cal{L}}$. The POVM consists of three
rank-two elements
\begin{subequations}
\label{POVM2}
  \begin{eqnarray}
\!\!\!\!\!\!\!\!\!\!{F}_{v,z}&=&\sqrt{\frac{2}{3}}(\ketbra{000}+\ketbra{111}) \\
\!\!\!\!\!\!\!\!\!\!{F}_{v,x}&=&\sqrt{\frac{2}{3}}(\ketbra{+++}+\ketbra{---})\\
\!\!\!\!\!\!\!\!\!\!{F}_{v,y}&=&\sqrt{\frac{2}{3}}(\ketbra{i,i,i}+\ketbra{-\!i,-\!i,-\!i}),
\end{eqnarray}
\end{subequations}
which extend those in Eq.~(\ref{eqn:F1D}) to three virtual qubits.
Note that  $|0/1\rangle$,
$\ket{\pm}\equiv(\ket{0}\pm\ket{1})/\sqrt{2}$ and $\ket{\pm i}\equiv
(\ket{0}\pm i\ket{1})/\sqrt{2}$ are eigenstates of Pauli operators
$Z$, $X$ and $Y$, respectively. Physically, $F_{v,a}$ is
proportional to a projector onto the two-dimensional subspace
spanned by the $S_a=\pm 3/2$ states, i.e., $\ketbra{S_a=\pm3/2}$. We
have simply used the three-virtual-qubit representation, and it will
be useful for our proof. The above POVM elements obey the relation
$\sum_{\nu \in \{x,y,z\}}F^\dagger_{v,\nu} F_{v,\nu} = P_{S,v}$,
i.e., project onto the symmetric subspace of three qubits,
equivalently, the identity in $S=3/2$ Hilbert space, as required.
The outcome of the POVM at any site $v$ is random,  $x$,
$y$ or $z$, and it can be correlated with the outcomes at other
sites due to correlations in the AKLT state. As we demonstrate
below, the resulting quantum state, dependent on the random POVM
outcomes ${\cal{A}}=\{a_v, v\in V({\cal{L}})\}$,
\begin{equation}
  \label{CA}
  |\Psi({\cal{A}})\rangle = \bigotimes_{v \in V({\cal{L}})} \!\! F_{v,a_v}\, |\Phi_{\rm AKLT}
\rangle= \bigotimes_{v \in V({\cal{L}})} \!\! F_{v,a_v} \bigotimes_{e \in
E({\cal{L}})} |\phi\rangle_e
\end{equation}
is equivalent under local unitary transformations to an encoded graph state
$\overline{|G({\cal{A}})\rangle}$.  The graph $G({\cal{A}})$ determines the
corresponding graph state, and we show that it is constructed from the
honeycomb lattice graph by applying the following two rules, given
${\cal{A}}$:

\begin{itemize}
\item[R1]{(Edge contraction): Contract all edges $e \in E({\cal{L}})$ that connect sites with the same POVM outcome.}
\item[R2]{(Mod 2 edge deletion): In the resultant multi-graph, delete
all edges of even multiplicity and convert all edges of odd
multiplicity into conventional edges of multiplicity 1.}
\end{itemize}
These two rules are illustrated in Fig.~\ref{merge}. A set of sites
in ${\cal{L}}$ that is contracted into a single vertex of
$G({\cal{A}})$ by the above rule R1 is called a {\em{domain}}, which
we have already encountered in the reduction of 1D AKLT state. Each
domain supports a single encoded qubit. The stabilizer generators
and the encoded operators for the resulting codes are summarized in
Table~\ref{tbl:coding2}. Below we demonstrate the post-POVM state
$|\Psi({\cal A})\rangle$ is a graph state and justify rules R1 and
R2 with simple examples.

\begin{figure}
 \includegraphics[width=8cm]{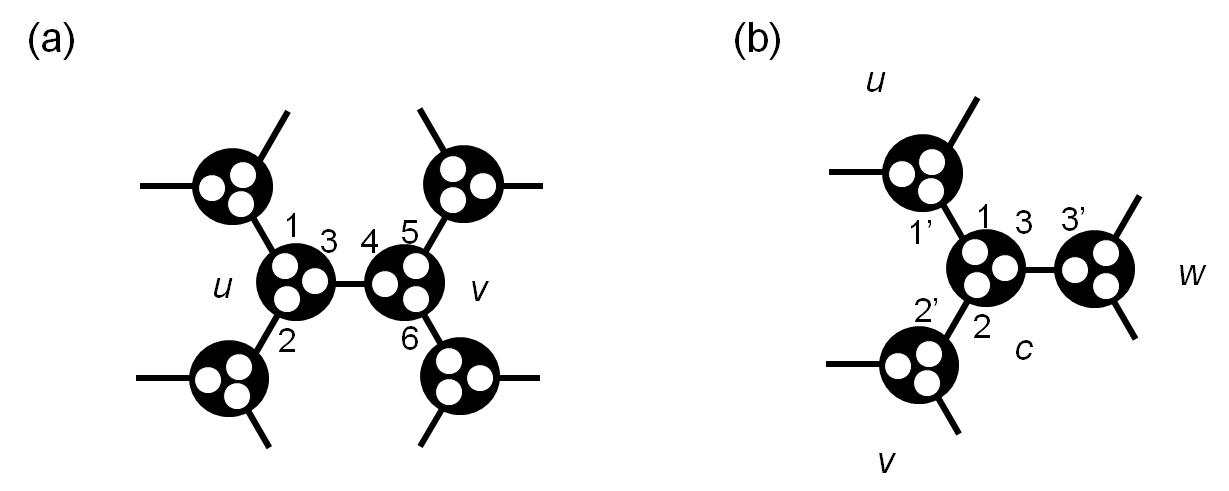}
  \caption{\label{fig:encodingKv}
  Illustrations of the encoding and the cluster graph stabilizer. (a) If two neighboring sites $u$ and $v$ share the same POVM
  outcome, e.g., $a_u=a_v=z$, then collectively these two sites form a logical qubit. More generally, if a set of connected sites
  share the same POVM outcome, then these sites effective encode a logical qubit. (b) An example of four sites $u$, $v$, $w$, and $c$
  with POVM outcomes $a_c=z$, and $a_u=a_v=a_w=x$ is used to illustrate the stabilizer generator $K_c=\pm \bar{X}_c \bar{Z}_u \bar{Z}_v\bar{Z}_w$.
  }
\end{figure}

Rule 1: Merging of sites. Physically, this rule derives from the
antiferromagnetic property of the AKLT state: neighboring spin-3/2 particles
must not have the same $S_a=3/2$ (or -3/2)
configuration~\cite{AKLT,AKLT2,AKLT3}. Hence, after the projection onto
$S_a=\pm3/2$ subspace by the POVM, the configurations for all sites inside a
domain can only be $\ket{3/2,-3/2,\dots}$ or $\ket{-3/2,3/2,\dots}$, and
intuitively, these form the basis of a single qubit. This encoding of a qubit
can also be understood in terms of the stabilizer.
 Consider the case where two neighboring POVMs yield the same
outcome, say $z$; see Fig.~\ref{fig:encodingKv}a. As a result of the
projections $F_{u,z}$ and $F_{v,z}$ (with $u=\{1,2,3\}$ and
$v=\{4,5,6\}$ each containing three virtual qubits), the operators
$Z_1Z_2$, $Z_2Z_3$, and $Z_4Z_5$, $Z_5 Z_6$ become stabilizer
generators of the post-POVM state $|\Psi({\cal{A}})\rangle$. In
addition, the stabilizer $-Z_3Z_4$ of the singlet state
$|\phi\rangle_{34}$ commutes with the projection $F_{u,z}\otimes
F_{v,z}$, and thus remains a stabilizer element for
$|\Psi({\cal{A}})\rangle$. In summary, the stabilizer generators are
$Z_1Z_2, Z_2Z_3, -Z_3Z_4, Z_4Z_5, Z_5Z_6$, giving rise to a single
encoded qubit
$$
  \alpha |(000)_u(111)_v\rangle + \beta |(111)_u(000)_v\rangle,
$$
which is supported by the two sites $u$ and $v$ jointly. We observe here the
antiferromagnetic ordering \cite{AKLT,AKLT2,AKLT3} among groups of three
virtual qubits. To reduce the support of this logical qubit to an individual
site of ${\cal{L}}$, a measurement in the basis $\{|(000)_v\rangle \pm
|(111)_v\rangle\}$ is performed. The resulting state is $\alpha
|(000)_u\rangle \pm \beta |(111)_u\rangle$, with the sign ``$\pm$'' known from
the measurement outcome. This is the proper encoding for a domain consisting
of a single site. Domains of more than two sites are thereby reduced to a
single site in the same manner.

To see that the state $|\Psi({\cal{A}})\rangle$ is indeed equivalent
under local unitary transformations to the encoded graph state
$\overline{|G({\cal{A}})\rangle}$, we consider the example of four
domains $c$, $u$, $v$, $w$, each consisting of a single site of
${\cal{L}}$, where the POVM outcome is $z$ on the central domain $c$
and $x$ on all lateral domains $u$, $v$ and $w$; see
Fig.~\ref{fig:encodingKv}b. By similar arguments as above, the
operator ${\cal O}\equiv -X_1 X_{1'}X_2 X_{2'}X_3 X_{3'}$ is in the
stabilizer of $|\Psi({\cal{A}})\rangle$. Using the encoding in
Table~\ref{tbl:coding2}, i.e., with the encoded Pauli operators
$\overline{X}_c=Z_1Z_2Z_3$, $\overline{Z}_u=\pm X_{1'}$,
$\overline{Z}_v=\pm X_{2'}$, and $\overline{Z}_w=\pm X_{3'}$, we
find that ${\cal O}=\pm\overline{X}_c
\overline{Z}_u\overline{Z}_v\overline{Z}_w$ which is (up to a
possible sign of convention) one of the stabilizer generators
defining the graph state. Intuitively, we see that each edge from an
outer domain contributes to an encoded $\overline{Z}$ from that
domain and the operator restricted at the center domain is clearly
not an identity nor an encoded $\overline{Z}$ as its POVM outcome
differs from the outer ones. This gives rise to a stabilizer
generator  local-unitarily equivalent to the one given in
Eq.~(\ref{eqn:stabilizerGen}).

Rule 2: Mod 2 edge deletion. By the above construction, if two domains $u$,
$v$ are connected by an edge of multiplicity $m$, the inferred graph state
stabilizer generators will contain factors of
$\overline{X}_u{\overline{Z}_v}^m$ or $\overline{X}_v{\overline{Z}_u}^m$. We
observe that $Z^2 = I$, from which Rule 2 follows.

Generalizing the above ideas, it is straightforward to rigorously
prove that for any POVM outcomes for any ${\cal{A}}$, the state
$|\Psi({\cal{A}})\rangle$ is local equivalent to an encoded graph
state $\overline{|G({\cal{A}})\rangle}$; see below. We shall denote
by $|G({\cal A})\rangle$ the graph state after reducing multiple
sites in every domain to a single site, i.e., to the proper qubit
encoding by domains, as the graph remains the same.

\section{From the AKLT state to graph states: general proof}
\label{sec:generalproof} Let us recall the POVM to be performed on
all sites can be rewritten as:
\begin{equation}
  \label{POVM2a}
  \begin{array}{rcl}
    F_{v,z} &=& \displaystyle{\sqrt{\frac{2}{3}}\frac{I_{12}+Z_1Z_2}{2}\frac{I_{23}+Z_2Z_3}{2}},\\
    F_{v,x} &=& \displaystyle{\sqrt{\frac{2}{3}}\frac{I_{12}+X_1X_2}{2}\frac{I_{23}+X_2X_3}{2}},\\
    F_{v,y} &=&
    \displaystyle{\sqrt{\frac{2}{3}}\frac{I_{12}+Y_1Y_2}{2}\frac{I_{23}+Y_2Y_3}{2}}.
  \end{array}
\end{equation}

It turns out that, for any ${\cal{A}}$, the state
$|\Psi({\cal{A}})\rangle$ is local equivalent to an encoded graph
state $\overline{|G({\cal{A}})\rangle}$, with the graph
$G({\cal{A}})$ constructed as follows. An edge $(v,w) \in
E({\cal{L}})$ is called internal iff at the sites $v$ and $w$ the
local POVM has resulted in the same outcome. The graph
$G({\cal{A}})$ is obtained from the lattice graph ${\cal{L}}$ by (1)
contracting all internal edges, and, in the resultant multi-graph,
(2a) deleting all edges of even multiplicity and (2b) converting all
edges of odd multiplicity into conventional edges of multiplicity 1.
See Fig.2 for illustration.

In step (1) of the above procedure, several sites of ${\cal{L}}$ are
merged into a single composite object ${\cal{C}} \in
V(G({\cal{A}}))$. Each such ${\cal{C}}$ is both a {\em{vertex}} in
the graph $G({\cal{A}})$ and a connected set of same-type sites of
${\cal{L}}$, i.e., a {\em{domain}}. Physically, in a domain of type
$a$, we have antiferromagnetic order along the $\pm a$-direction,
because two neighboring spins never have the same $S_a=3/2$ (or
-3/2) in the AKLT state~\cite{AKLT}. The state of the domain
contains only two configurations w.r.t. the quantization axis $a$:
$|+3/2,-3/2,+3/2,\dots\rangle$ and $|-3/2,3/2,-3/2,\dots\rangle$.
Thus, it is effectively one qubit.

The outlined construction leads to one of our main results:

\begin{Thm}
  \label{Creduct}
  For any ${\cal{A}}$ that specifies all outcomes of POVMs on ${\cal L}$,
  quantum computation by local spin-3/2 measurements on the state $|\Psi({\cal{A}})\rangle$ can efficiently simulate quantum computation by local spin-1/2 measurement on the graph state $|G({\cal{A}})\rangle$.
\end{Thm}

Thus, the computational power of the AKLT state,  as harnessed by
the POVMs Eq.~(\ref{POVM2a}), hinges on the connectivity properties
of $G({\cal{A}})$. If, for typical sets ${\cal{A}}$ of POVM
outcomes, the graph state $|G({\cal{A}})\rangle$ is computationally
universal then so is the AKLT state.

 The  proof proceeds in three steps. First
we show that every domain ${\cal{C}} \in V(G({\cal{A}}))$ gives rise
to one encoded qubit. Second, we show that $|\Psi({\cal{A}})\rangle$
is, up to local encoded unitaries, equivalent to the encoded graph
state $\overline{|G({\cal{A}})\rangle}$. Third, we show that the
encoding can be unraveled by local spin-3/2 measurements.

{\em{Step 1: Encoding.}} Consider a domain ${\cal{C}} \subset
V({\cal{L}})$. That is, on all sites $v \in {\cal{C}}$ the same POVM
outcome $a \in \{x,y,z\}$ was obtained. ${\cal{C}}$ contains
$3|{\cal{C}}|$ qubits. The projections $F_{v,a}$ on all $v \in
{\cal{C}}$ enforce $2|{\cal{C}}|$ stabilizer generators, c.f.
Eq.~(\ref{POVM2}). Furthermore, choose a tree ${\cal{T}}$ among the
set of  edges with both endpoints in the domain ${\cal{C}}$. Each
edge $(u,v) \in {\cal{T}}$ contributes a stabilizer generator
$-\sigma_a^{(u)}\sigma_a^{(v)}$ to the product of Bell states
$\bigotimes_{e \in E({\cal{L}})} |\phi\rangle_e$. These stabilizers
commute with the local POVMs (\ref{POVM2a}) and therefore are also
stabilizer generators for $|\Psi({\cal{A}})\rangle$. Since
$|{\cal{T}}| = |{\cal{C}}|-1$, in total there are $3|{\cal{C}}|-1$
stabilizer generators with support only in ${\cal{C}}$, acting on
$3|{\cal{C}}|$ qubits. They give rise to one encoded qubit.

While the stabilizer generators for our code follow from the
construction, there is freedom in choosing the encoded Pauli
operators. Table~\ref{tbl:coding2} shows one such choice of
encoding.

\begin{table}
  \begin{tabular}{l|r|r|r}
    \parbox{1.4cm}{POVM outcome\vspace{1mm}} & \multicolumn{1}{c|}{$z$} & \multicolumn{1}{c|}{$x$} & \multicolumn{1}{c}{$y$} \\ \hline
    \parbox{1.6cm}{stabilizer generator} & $\lambda_{i}\lambda_{i+1} Z_iZ_{i+1}$, & $\lambda_{i}\lambda_{i+1} X_iX_{i+1}$ & $\lambda_{i}\lambda_{i+1}Y_iY_{i+1}$ \\
    $\overline{X}$ & $\bigotimes_{j = 1}^{3|{\cal{C}}|} X_j$ & $\bigotimes_{j = 1}^{3|{\cal{C}}|} Z_j$ & $\bigotimes_{j = 1}^{3|{\cal{C}}|} Z_j$ \\
    $\overline{Z}$ & $\lambda_i Z_i$ & $\lambda_i X_i$ & $\lambda_i Y_i$
  \end{tabular}
 \caption{\label{tbl:coding2} The dependence of stabilizers and encodings for the
 random graphs
  on the local POVM outcome. $|{\cal C}|$ denotes the total number of virtual qubits contained in a
  domain. In the first line, $i = 1\, ..\, 3|{\cal{C}}|-1$, and in the third line
  $i = 1\,..\, 3|{\cal{C}}|$. The honeycomb lattice ${\cal{L}}$ is bicolorable or bipartite and all sites can be
  divided into either $A$ or $B$ sublattice, $V({\cal{L}}) = A \cup B$.
  Then, one choice of the sign is $\lambda_i = 1$ if the virtual qubit $i \in v \in A$ and  $\lambda_i = -1$ if
   $i \in v' \in B$.}
\end{table}

{\em{Step 2:}} We show that $|\Psi({\cal{A}})\rangle$ is an encoded
graph state. Consider a central vertex ${\cal{C}}_c \in
V(G({\cal{A}}))$ and all its neighboring vertices ${\cal{C}}_\mu \in
V(G({\cal{A}}))$. Denote the POVM outcome for all ${\cal{L}}$-sites
$v \in {\cal{C}}_c, {\cal{C}}_\mu$ by $a_c$ and $a_\mu$,
respectively. Denote by $E_\mu$ the set of ${\cal{L}}$-edges that
run between ${\cal{C}}_c$ and ${\cal{C}}_\mu$. Denote by $E_c$ the
set of ${\cal{L}}$-edges internal to ${\cal{C}}_c$. Denote by $C_c$
the set of all qubits in ${\cal{C}}_c$, and by $C_\mu$ the set of
all qubits in ${\cal{C}}_\mu$. (Recall that there are 3 qubit
locations per ${\cal{L}}$-vertex $v \in {\cal{C}}_c,{\cal{C}}_\mu$.)
We first consider the stabilizer of the state $\bigotimes_{e \in
E({\cal{L}})}|\phi\rangle_e$. For any $\mu$ and any edge $e \in
E_\mu$, let $u(e)$ [$v(e)$] be the endpoint of $e$ in $C_\mu$
[$C_c$]. Then, for all $\mu$ and all $e \in E_\mu$ the Pauli
operators $-\sigma_{a_\mu}^{({u(e)})}\sigma_{a_\mu}^{(v(e))}$ are in
the stabilizer of $\bigotimes_{e \in E({\cal{L}})}|\phi\rangle_e$.
Choose $b \in \{x,y,z\}$ such that $b \neq a_c$, and let, for any
edge $e' \in E_c$, $v_1(e'), v_2(e') \in C_c$ be qubit locations
such that $e'=(v_1(e'), v_2(e'))$. Then, for all $e' \in E_c$,
$-\sigma_b^{({v_1(e')})}\sigma_b^{(v_2(e'))}$ is in the stabilizer
of $\bigotimes_{e \in E({\cal{L}})}|\phi\rangle_e$. Therefore, the
product of all these operators,
\begin{equation}
  O_{{\cal{C}}_c} = \pm \! \left(\bigotimes_{\mu}\bigotimes_{e \in E_\mu} \sigma_{a_\mu}^{(u(e))} \sigma_{a_\mu}^{(v(e))} \! \right)\!\! \left(\bigotimes_{e' \in E_c} \sigma_{b}^{(v_1(e'))} \sigma_{b}^{(v_2(e'))} \!\!\right)
\end{equation}
is also in the stabilizer of $\bigotimes_{e \in
E({\cal{L}})}|\phi\rangle_e$.

We now show that $O_{{\cal{C}}_c}$ commutes with the local POVMs and
is therefore also in the stabilizer of $|\Psi({\cal{A}})\rangle$.
First, consider the central domain ${\cal{C}}_c$. The operator
$O_{{\cal{C}}_c}$ acts non-trivially on every qubit in $C_c$,
$O_{{\cal{C}}_c}|_l \neq I_l$ for all qubits $l \in C_c$.
Furthermore, for all qubits $l \in C_c$,
$O_{{\cal{C}}_c}\left|_l\right. \neq \sigma_{a_c}^{(l)}$. Namely, if
$l \in C_c$ is connected by an edge $e \in E_\mu$ to
${\cal{C}}_\mu$, for some $\mu$, then
$O_{{\cal{C}}_c}\left|_l\right. = \sigma_{a_\mu}^{(l)}  \neq
\sigma_{a_c}^{(l)}$ (for all $\mu$, $a_\mu \neq a_c$ by construction
of $G({\cal{A}})$). Or, if $l \in C_c$ is the endpoint of an
internal edge $e' \in E_c$ then $O_{{\cal{C}}_c}\left|_l\right. =
\sigma_b^{(l)}  \neq \sigma_{a_c}^{(l)}$ ($a_c \neq b$ by above
choice). Therefore, for any $i,j \in C_c$, $O_{{\cal{C}}_c}$
anticommutes with $\sigma_{a_c}^{(i)}$ and $\sigma_{a_c}^{(j)}$, and
thus commutes with all $\sigma_{a_c}^{(i)}\sigma_{a_c}^{(j)}$. Thus,
$O_{{\cal{C}}_c}$ commutes with the local POVMs Eq.~(\ref{POVM2a})
on all $v \in {\cal{C}}_c$.

Second, consider the neighboring  domains ${\cal{C}}_\mu$.
$O_{{\cal{C}}_c}\left|_{{\cal{C}}_\mu}\right. =
\bigotimes_{j}\sigma_{a_\mu}^{(j)}$ by construction.
$O_{{\cal{C}}_0}$ thus commutes with the local POVMs $F_{v,a_\mu}$
for all $v \in {\cal{C}}_\mu$ and for all $\mu$.

Therefore, $O_{{\cal{C}}_c}$ is in the  stabilizer of
$|\Psi({\cal{A}})\rangle$. Therefore, $O_{{\cal{C}}_c}$ is an
encoded operator w.r.t. the code in Table~\ref{tbl:coding2}, and we
need to figure out which one. (1) Central vertex ${\cal{C}}_c$:
$\left. O_{{\cal{C}}_c}|_{C_c}\right.$ is an encoded operator on
$C_c$, $\left. O_{{\cal{C}}_c}|_{C_c}\right. \in \{\pm \overline{I},
\pm \overline{X}, \pm \overline{Y}, \pm \overline{Z}\}$. Since
$\left. O_{{\cal{C}}_c}|_l\right. \neq \sigma_{a_c}^{(l)}$ for any
$l \in {\cal{C}}_c$, by Table~\ref{tbl:coding2}, $\left.
O_{{\cal{C}}_c}|_{C_c}\right. \neq \pm \overline{I}, \pm
\overline{Z}$. Thus, $\left. O_{{\cal{C}}_c}|_{C_c}\right. \in \{\pm
\overline{X}, \pm \overline{Y}\}$. (2) Neighboring vertices
${\cal{C}}_\mu$: By Table~\ref{tbl:coding2}, $\sigma_{a_\mu}^{(l)} =
\pm \overline{Z}$, for any $l \in C_\mu$. Thus, $\left.
O_{{\cal{C}}_c}|_{C_\mu}\right. = \pm \overline{Z}^{|E_\mu|}$. Now
observe that $Z^2 = I$, and that, this justifies the above
prescription in constructing  the graph $G({\cal{A}})$. Using the
adjacency matrix $A_{G({\cal{A}})}$, we have $|E_\mu| \mod 2 =
[A_{G({\cal{A}})}]_{c,\mu}$ and hence $\left.
O_{{\cal{C}}_c}|_{C_\mu}\right. =\pm
\overline{Z}^{[A_{G({\cal{A}})}]_{c,\mu}}$.

Thus, finally, for all ${\cal{C}}_c \in V(G({\cal{A}}))$,
\begin{equation}
  \label{CSstab}
  O_{{\cal{C}}_c} \in \left\{ \pm \overline{R}_{C_c} \bigotimes_{{\cal{C}}_\mu \in V(G({\cal{A}}))}  \overline{Z}_{{\cal{C}}_\mu}^{[A_{G({\cal{A}})}]_{c,\mu}},\; \mbox{with } R = X,Y \right\}
\end{equation}
This is, up to conjugation by one of the  local encoded gates
$\overline{I}_{{\cal{C}}_c}, \overline{Z}_{{\cal{C}}_c},
\exp\big(\pm i\pi/4\, \overline{Z}_{{\cal{C}}_c}\big)$, a stabilizer
generator for the encoded graph state
$\overline{|G({\cal{A}})\rangle}$. The code stabilizers in
Table~\ref{tbl:coding2}
 and the stabilizer operators in Eq.~(\ref{CSstab})
together define the state $|\Psi({\cal{A}})\rangle$ uniquely.
$|\Psi({\cal{A}})\rangle$ is, up to the action of local encoded
phase gates, an encoded graph state $|G({\cal{A}})\rangle$.

{\em{Step 3: Decoding of the code.}}  We show that any domain
${\cal{C}} \in V(G({\cal{A}}))$ can be reduced to a single
elementary site $w \in V({\cal{L}})$ by local measurement on all
other sites $v \in {\cal{C}}$, $v \neq w$. For any such $v$, choose
the measurement basis ${\cal{B}}_a$, $a \in \{x,y,z\}$, as follows
\begin{equation}
  \label{Meas}
  \begin{array}{rcl}
    {\cal{B}}_x &=&\left\{{(|+++\rangle \pm |---\rangle)}/{\sqrt{2}} \right\}, \\
    {\cal{B}}_y &=& \left\{{(|i,i,i\rangle \pm|-i,-i,-i\rangle)}/{\sqrt{2}}\right\}, \\
    {\cal{B}}_z&=&\left\{{(|000\rangle \pm |111\rangle)}/{\sqrt{2}}\right\}.
  \end{array}
\end{equation}
These measurements map the symmetric subspace of the three-qubit
states into itself and  they can therefore be performed on the
physical spin 3/2 systems.

Denote by ${\cal{S}}_{\cal{C}}$ and ${\cal{S}}_{{\cal{C}}\backslash
v}$ the code stabilizer on the domain ${\cal{C}} \in
V(G({\cal{A}}))$ and on the reduced domain ${\cal{C}}\backslash v$,
respectively. Using standard stabilizer techniques~\cite{Stabilizer}
it can be shown that the measurement Eq.~(\ref{Meas}) has the
following effect on the encoding
\begin{equation}
    {\cal{S}}_{\cal{C}} \longrightarrow  {\cal{S}}_{{\cal{C}}\backslash v},\;
    \overline{X}_{\cal{C}} \longrightarrow \pm \overline{X}_{{\cal{C}}\backslash v},\;
    \overline{Z}_{\cal{C}} \longrightarrow \overline{Z}_{{\cal{C}}\backslash v}.
\end{equation}
The measurement (\ref{Meas}) thus removes from ${\cal{C}}$ by  one
lattice site $v \in V({\cal{L}})$. We repeat the procedure until
only one site, $w$, remains in ${\cal{C}}$, for each ${\cal{C}} \in
V(G({\cal{A}}))$. In this way, ${\cal{S}}_{\cal{C}} \longrightarrow
{\cal{S}}_{\{w\}}$, $\overline{X}_{\cal{C}} \longrightarrow \pm
\overline{X}_{\{w\}}$, $\overline{Z}_{\cal{C}}  \longrightarrow
\overline{Z}_{\{w\}}$. Thus, $|\Psi(A)\rangle \longrightarrow
\overline{U}_\text{loc} \overline{|G(A)\rangle}=: |G(A)\rangle$,
where $U_\text{loc}$ is a local unitary, and the encoding in
Table~\ref{tbl:coding2} has now shrunk to one site of ${\cal{L}}$
per encoded qubit, i.e. to three auxiliary qubits.

To complete the computation,  the remaining encoded qubits are
measured individually. Again, the measurement of an encoded qubit on
a site $w \in {\cal{L}}$ is an operation on the symmetric subspace
of three auxiliary qubits at $w$, and can thus be realized as a
measurement on the equivalent physical spin 3/2. $\Box$

\section{Random graph states, percolation and 2D cluster states}
\label{sec:RandomGraph}

Whether or not typical graph states $|G({\cal{A}})\rangle$ are
universal resources hinges solely on the connectivity properties of
$G({\cal{A}})$, and is thus a percolation problem~\cite{Perc}.  We
test whether, for typical graphs $G({\cal{A}})$,
\begin{enumerate}
  \item[C1]{\label{C1}The size of the largest domain scales at most logarithmically with the total number of sites $|V({\cal{L}})|$.}
  \item[C2]{\label{C2}Let $S \subset {\cal{L}}$ be a rectangle of size $l \times 2l$ ($2l \times l$). Then, a path through $G|_S$  from the left to the right (top to bottom) exists with probability approaching 1 in the limit of large ${\cal{L}}$.}
\end{enumerate}
Note that Condition~C1 is obeyed whenever the domains are
{\em{microscopic}}, i.e., their size distribution is independent of
$|{\cal{L}}|$ in the limit of large ${\cal{L}}$. Then, the size of
the largest domain scales logarithmically in $|{\cal{L}}|$
\cite{Perc}. Condition~C2 ensures that the system is in the
percolating phase.

Together with planarity, which holds for all graphs $G({\cal{A}})$
by construction,  the conditions C1 and C2 are sufficient for the
reduction of the random graph state to a standard universal cluster
state. The proof given below  extends a similar result already
established for site percolation on a square lattice~\cite{BPerc}.
The physical intuition comes from percolation theory. In the
percolating (or supercritical) phase, the spanning cluster contains
a subgraph which is topologically equivalent to a coarse-grained
two-dimensional lattice structure. This subgraph can be carved out
and subsequently cleaned off all imperfections by local Pauli
measurements, leading to a perfect two-dimensional lattice.

\subsection{Reduction of $|G({\cal{A}})\rangle$ to a 2D cluster state
above the percolation threshold}\label{2Dreduc}

We define the distance $\text{\em{dist}}_{\cal{L}}(v,w)$ between two
vertices $v, w \in V({\cal{L}})$ as the minimum number of edges on a
path between $u$ and $v$, and consider two further properties of
graphs $G({\cal{A}})$:
\begin{enumerate}
\item[{C1$'$}]{$G({\cal{A}})$ can be embedded in ${\cal{L}}$ such that the maximum distance between the endpoints of an edge in $E(G)$ scales at most logarithmically in $|{\cal{L}}|$,}
\item[{C2$'$}]{Let $S \subset {\cal{L}}$ be a rectangle of size $l \times 5l$ ($5l \times l$). Then, a path through $G|_S$  from the left to the right (top to bottom) exists with probability approaching 1 in the limit of large ${\cal{L}}$.}
\end{enumerate}
\begin{Lemma}
\label{L1} $G({\cal{A}})$ is planar for all POVM outcomes
${\cal{A}}$. Property C1 implies property C1$'$, and property C2
implies property C2\,$'$.
\end{Lemma}

{\em{Proof of Lemma~\ref{L1}.}} Planarity: $G({\cal{A}})$ is
obtained from the honeycomb lattice ${\cal{L}}$, which is planar, by
the graph rules R1 and R2. They only perform edge deletion and edge
contraction, which preserve the planarity of ${\cal{L}}$.
$G({\cal{A}}$ is thus planar for all ${\cal{A}}$.

Property~C1$'$: For any domain $d \subset V({\cal{L}})$, place the
corresponding vertex $v(d) \in V(G({\cal{A}}))$ inside $d$ such that
the distance $r(d):=\max_{w\in d}\text{dist}_{\cal{L}}(w,v)$ is
minimized. Then, $r(d)\leq |d|/2$. Now, consider two vertices
$v(d_1), v(d_2) \in V(G)$ connected by an edge $e \in E(G)$. Then,
the domains $d_1$, $d_2$ are connected by a single edge in
$E({\cal{L}})$. Thus, for any pair $(v(d_1), v(d_2))$ of vertices in
$G$ connected by an edge in $E(G)$, $\text{dist}_{\cal{L}}(v(d_1),
v(d_2)) \leq (|d_1|+|d_2|)/2 +1$. By property C1, this length scales
at most logarithmically in $|{\cal{L}}|$.

\begin{figure}
\begin{center}
   \includegraphics[width=5.5cm]{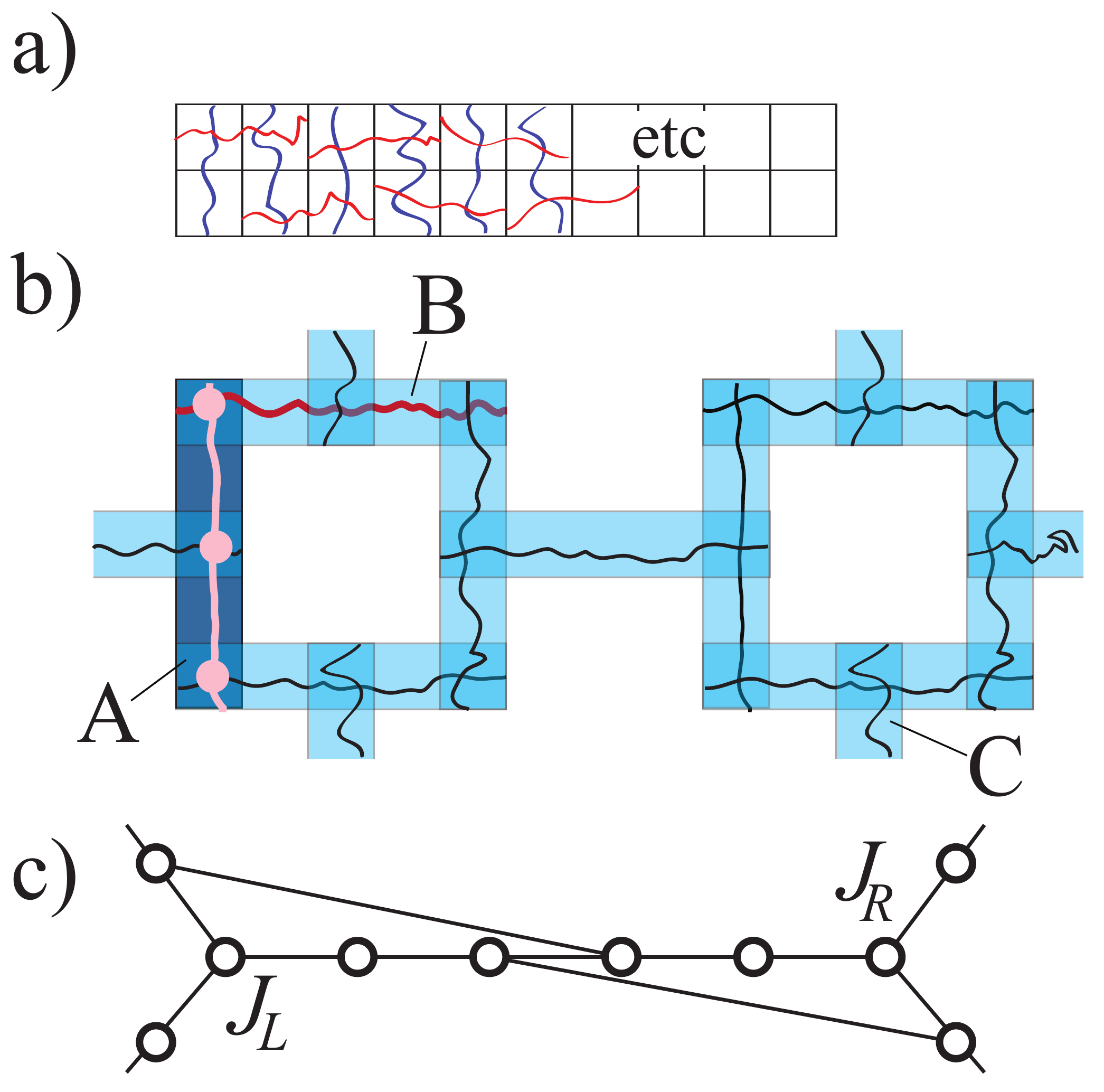}
\caption{\label{P1}(a) Constructing a rectangle with a traversing
path of aspect ratio 1:5, from such rectangles with aspect ratio
1:2, 2:1. (b) Overlapping rectangles of size $L \times 5L$,
$5L\times L$. The union of their traversing paths yields the net
${\cal{P}}$. (c) Pair of non-separated junctions.}
\end{center}
\end{figure}

Property~C2$'$: See Fig.~\ref{P1}a.\medskip

\begin{Lemma}
\label{SfU} Consider a planar graph $G({\cal{A}}_{\cal{L}})$
embedded into the lattice ${\cal{L}}$ of size $\Lambda \times
\Lambda$, satisfying the properties C1$'$ and C2\,$'$. Then, the
graph state $|G({\cal{A}}_{\cal{L}})\rangle$ can be converted by
local measurements to a two-dimensional cluster state of size
$\Lambda' \times \Lambda'$, with $\Lambda' \sim \Lambda/\log
\Lambda$.
\end{Lemma}
Both lemmata combined give the desired result:
\begin{Thm}
Consider an AKLT state on a honeycomb lattice ${\cal{L}}$ converted
into a random graph state $|\overline{G({\cal{A}})}\rangle$ by the
POVM (\ref{POVM2}). If for typical POVM outcomes ${\cal{A}}$ the
corresponding graph $G({\cal{A}})$ satisfies the conditions C1 and
C2, then the AKLT state is a universal resource for MBQC.
Furthermore, the computation requires at most a poly-logarithmic
overhead compared to cluster states.
\end{Thm}

{\em{Remark:}} The polylog bound to the overhead comes from bounding
the average domain size by the maximum domain size, for technical
reasons. The true overhead is expected to be constant.

That the conditions C1 and C2 are obeyed in the typical case remains
to be demonstrated. We show this numerically, as reported in the
next section.\medskip

{\em{Proof of Lemma~\ref{SfU} - main tool.}} We show that a  graph
state $|G\rangle$ satisfying the assumptions of Lemma~\ref{SfU} can
be reduced to the cluster state on a two-dimensional square grid, by
local Pauli measurement on a subset of its qubits. The 2D cluster
state is already known to be universal \cite{Oneway}. Specifically,
we use the following rules \cite{Hein} for the manipulation of graph
states
\begin{subequations}
  \label{GraphRules}
  \begin{align}
  \label{R1}
  \parbox{5cm}{\includegraphics[width=5cm]{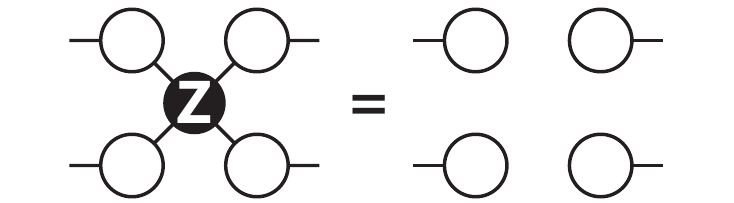}},\\
  \label{R2}
  \parbox{5cm}{\includegraphics[width=5cm]{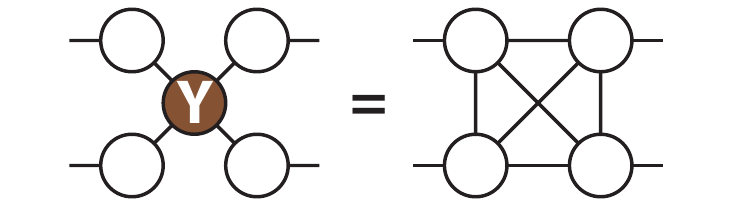}},\\
   \label{R3}
  \parbox{5cm}{\includegraphics[width=5cm]{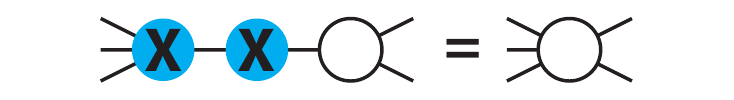}}.
  \end{align}
\end{subequations}
Rule (\ref{R1}): The effect of a $\sigma_z$-measurement at vertex $a$ on the interaction graph is to remove $a$ and all edges ending in $a$. \\
Rule (\ref{R2}): The effect of a $\sigma_y$-measurement at vertex $a$ on the interaction graph is to invert all edges in the neighborhood $N(a)$ of $a$, and to remove $a$ and all edges adjacent to $a$.\\
Rule (\ref{R3}): Consider three qubits on a line, where the middle
qubit has exactly two neighbors. When the left and the middle qubit
are measured in the $\sigma_x$-basis, the interaction graph changes
as follows: The right vertex inherits the neighbors of the left
vertex. The left and the middle vertex plus all edges adjacent to
them are deleted.

\begin{figure}
\begin{center}
  \includegraphics[width=7.5cm]{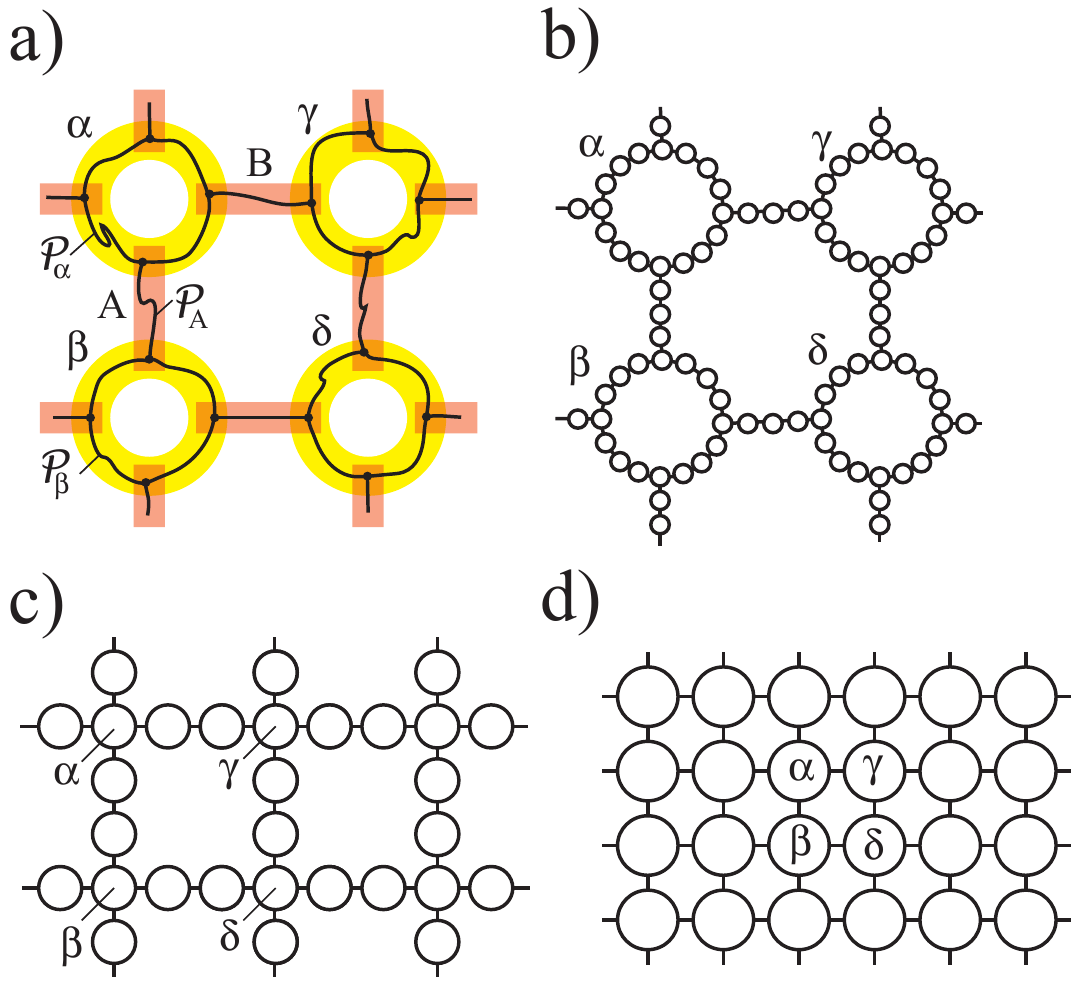}
  \caption{\label{perc}Transforming $|G({\cal{A}})\rangle$ into a 2D cluster state. (a) Macroscopic view: Regions $\alpha$, $\beta$, $A$ etc imposed on the graph $G({\cal{A}})$. (b) Graph with three-valent junctions. (c) Decorated 2D grid. (d) 2D grid for the cluster state.}
\end{center}
\end{figure}

{\em{Proof of Lemma~\ref{SfU} - Outline.}} We consider a graph $G$
with properties C1$'$, C2$'$. We impose a pattern of regions $A$,
$B$, $C$, .. of rectangular shape and size $l \times 5l$ and $5l
\times l$, for sufficiently large $l$, on the plane into which $G$
is embedded; See Fig.~\ref{P1}b. Due to the percolation property,
$G$ has a net-shaped subgraph ${\cal{P}}$, shown in
Fig.~\ref{perc}a. In the first step of the reduction all qubits in
$V(G)\backslash V({\cal{P}})$ are measured individually in the
$\sigma_z$-basis. The graph thereby created is close to the one
displayed in Fig.~\ref{perc}b. However, it may have additional edges
that cannot be removed by vertex deletion (\ref{R1}) alone. Such
edges are removed by a combination of the graph rules (\ref{R1}) and
(\ref{R2}). Then, in two further steps, the graph state of
Fig.~\ref{perc}b is converted to the 2D cluster state shown in
Fig.~\ref{perc}d.

{\em{Step 1.}} -  Conversion of the graph $G$ to the graph of
Fig.~\ref{perc}b by local operations. First we show that the net
${\cal{P}}$ of paths shown in Fig.~\ref{perc}a exists. Consider the
overlapping rectangles $A$ and $B$ in Fig.~\ref{P1}b. By Property
C2$'$, $A$ has a path ${\cal{P}}_A$ running from top to bottom, and
$B$ a path ${\cal{P}}_B$ crossing from left to right. Since $G$ is
planar, ${\cal{P}}_A$ and ${\cal{P}}_B$ must intersect in at least
one vertex. The net ${\cal{P}}$ is defined to be the union of all
such traversing paths (one per rectangle), with all ends removed
that do not affect connectedness. ${\cal{P}}$ is shown in
Fig.~\ref{perc}a.

Now, $G$ is converted to $G|_{V({\cal{P}})}$, by deleting all
vertices from $G$ which are not in $V({\cal{P}})$. Physically, the
corresponding graph state $|G|_{V({\cal{P}})}\rangle$ is obtained by
measuring the qubits at all vertices $v \in V(G)\backslash
V({\cal{P}})$ in the eigenbasis of $\sigma_z$, c.f. Eq.~(\ref{R1}).
After that, ideally, all junctions of paths in  $G|_{V({\cal{P}})}$
should be T-shaped, as in the graph of Fig.~\ref{perc}b. However, in
general they will not be. The quintessential (but not only)
obstruction is
$$
\includegraphics[width=2.5cm]{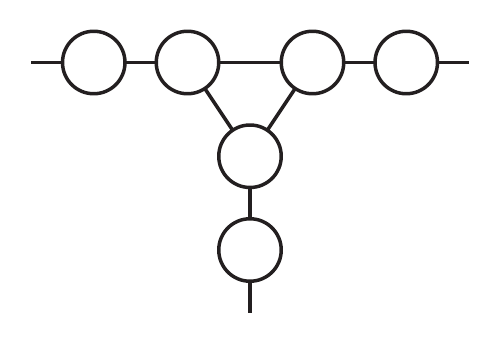}
$$
Note that no further edges can be removed by vertex deletion
(corresponding to $\sigma_z$-measurements), without disconnecting
the junction. Nonetheless, this obstruction is easily dealt with.
The above ring junction is converted into a T-junction by a single
measurement in the $\sigma_y$-basis,
\begin{equation}
\label{Yred}
\parbox{6.5cm}{\includegraphics[width=6.5cm]{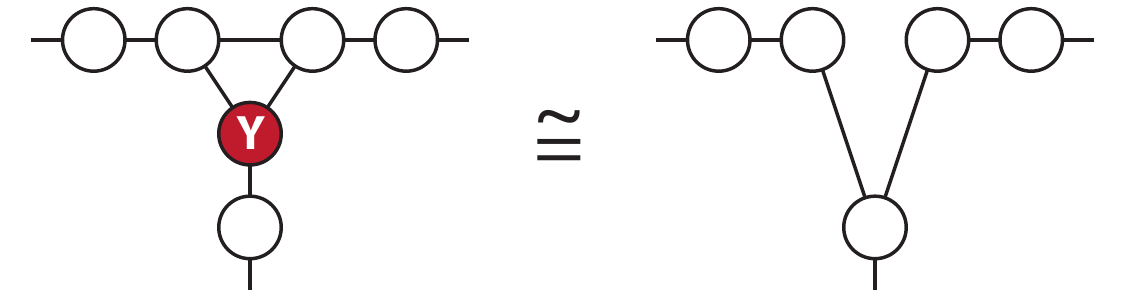}}
\end{equation}
However, we have to show that all possible obstructions to
T-junctions can be removed.

We begin with the wires, running from one junction $J_L \in
V(G|_{V({\cal{P}})})$ to another junction $J_R \in
V(G|_{V({\cal{P}})})$. They also have obstructions, for example
$$
\includegraphics[width=5cm]{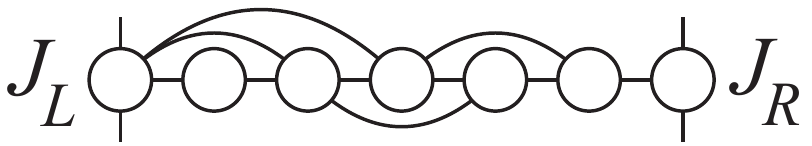}
$$
The first goal is to remove all obstructions within each wire. We
require that for a given wire, each belonging vertex, except $J_L$
and $J_R$, has a unique neighbor to the right and a unique neighbor
to the left in the wire, to which it is connected by an edge. $J_L$
($J_R$) only has a unique right but no left (a unique left but no
right) neighbor in the wire. The vertices are allowed to have edges
with vertices in other wires. Those will be removed later.

Note that the vertices in the wire are left-right ordered, by order
of appearance in the corresponding percolation path. Now, the
obstructions are removed from any given wire by the following
procedure.  Starting with $v=J_L$, take a vertex $v$ and identify
its rightmost neighbor in the wire, $w$. Delete all vertices between
$v$ and $w$. Now set $v:= w$, and repeat until $w=J_R$.

In this way, each vertex in the wire remains with a single right
neighbor in the wire. Therefore, the number of edges within the wire
equals the number of vertices minus 1. Therefore, each vertex in the
wire also has a unique left neighbor.

At this stage we remain with the obstructions at junctions. First we
show that we can treat the junctions individually, by choosing a
sufficiently large length scale $l$ for the size $l \times 5l$- ($5l
\times l$-) rectangles. Two neighboring junctions $J_L$, $J_R$ have
a distance of at least $l$, c.f. Fig.~\ref{P1}b. They are separated
if the configuration of edges shown in Fig.~\ref{P1}c does never
occur. It doesn't occur if $l> 2 |e|_{\text{max}}$, where
$|e|_{\text{max}}$ is the maximum distance of an edge in
$E(G({\cal{A}}))$. With C1$'$ it thus suffices to choose
\begin{equation}
\label{LLamb}
  l \sim log \Lambda.
\end{equation}
Now we discuss an individual junction $J$. By construction it joins
three wires, $W_L$, $W_C$ and $W_R$ say. The obstructions are three
sets of edges, $E_{LR}$, $E_{LC}$ and $E_{CR}$. They connect
vertices in $W_L$ with vertices $W_R$, $W_L$ and $W_C$, and $W_C$
and $W_R$, respectively. By the choice Eq.~(\ref{LLamb}) for $l$,
the obstructions at different junctions are well separated from each
other, and we can thus treat them individually.

First, we remove the obstructions $E_{LC}$ and $E_{CR}$, by the
following procedure. We approach the junction at $J$ on the wire
$W_C$. Denote by $v$ the first vertex in $W_C$ which is the endpoint
of an edge $e \in E_{LC}\cup E_{CR}$. By rule (\ref{R1}), delete all
vertices in $W_C$ between $v$ and $J$, excluding $v$ and $J$. Then,
there arise three cases. (1) $v$ is connected to a single vertex in
$W_L \cup W_R$. (2)  $v$ is  connected to exactly two vertices $w_1,
w_2 \in W_L \cup W_R$, and $w_1$, $w_2$ are neighbors in $W_L \cup
W_R$. (3) $v$ is connected to exactly two vertices $w_1, w_2 \in W_L
\cup W_R$, and $w_1, w_2$ are not neighbors in $W_L \cup W_R$; or
$v$ is connected to more than two vertices in  $W_L \cup W_R$.
Graphically, the cases look as follows (the obstructing edges
$E_{LR}$  are not relevant in the present sub-step, and are not
shown),
$$
\includegraphics[width=8cm]{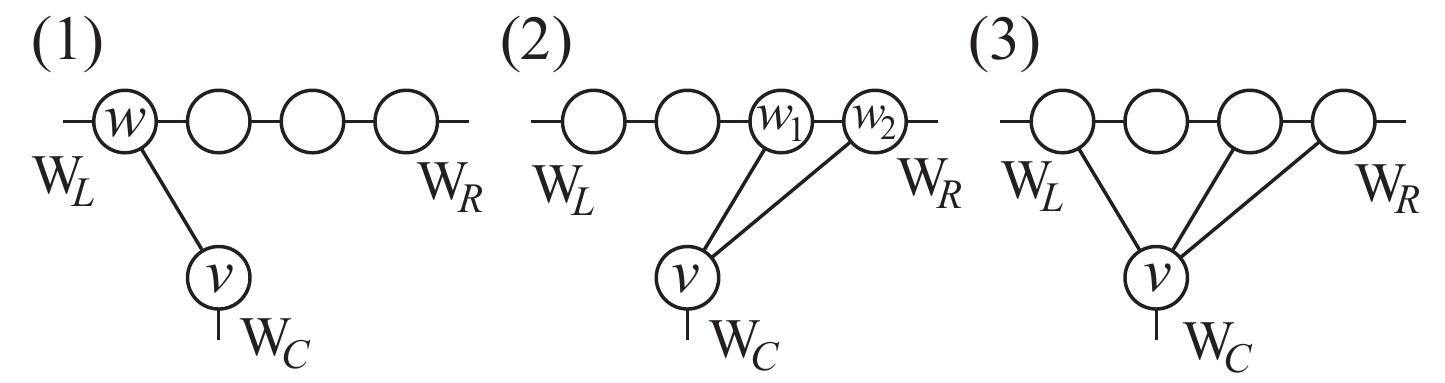}
$$
Case (1): The vertex $w$ is taken as the new junction center $J$,
and the edge $(v,w)$ is included into $W_C$. Thereby, the
obstructing edge sets $E_{LC}$ and $E_{CR}$ are removed. Case (2):
The qubit on vertex $v$ is measured in the $\sigma_y$-basis; c.f.
Eq.~(\ref{Yred}). Thereby, case (2) is reduced to case (1). Case
(3): Denote the leftmost (rightmost) neighbor of $v$ in $W_L \cup
W_R$ by $w_L$ ($w_R$). All vertices in $W_L \cup W_R$ between $w_L$
and $w_R$ are deleted, by $\sigma_z$-measurement on the
corresponding qubits. Thereby, case (3) is reduced to case (1).

In the above procedure, the center $J$ of the junction may have
shifted within $W_L \cup W_R$. Consequently, $W_L$ (to the left of
$J$), $W_R$ (to the right of $J$) and $E_{LR}$ are modified. Due to
the shift in the location of $J$, edges that were in $E_{LR}$ may
have become obstructing edges internal to $W_L$ or to $W_R$. They
are removed by re-running the previous procedure for removing
obstructing edges internal to the wires. There are no new edges in
$E_{LR}$.

By the above procedure, we have created a junction of three wires
$W_L$, $W_C$ and $W_R$ which are free of all obstructions except for
the set $E_{LR}$. These edges are now removed by approaching the new
junction center $J$ from the wire $W_L$ and repeating the previous
procedure.

{\em{Step 2}} - Creating the decorated lattice graph of
Fig.~\ref{perc}c. Consider a ring-shaped segment of the graph in
Fig.~\ref{perc}b, with the four belonging T-junctions $\alpha$,
$\beta$, $\gamma$ and $\delta$. The qubits on all vertices on the
path between $\alpha$ and $\beta$ are measured in the
$\sigma_z$-basis. The corresponding vertices are thereby removed,
c.f. Eq.~(\ref{R1}). Regarding the path between $\alpha$ and
$\delta$, we require that $\alpha$ and $\beta$ are not neighbors.
This can always be arranged by starting with a sufficiently large
scale $l$ for the $l \times 5l$-rectangles. If the the number of
vertices that lie on the path between $\alpha$ and $\delta$ is even
(but $>0$), then the qubit on the vertex next to $\alpha$ is
measured in the $\sigma_y$-basis. In this way, the number of
vertices between $\alpha$ and $\delta$ becomes odd, c.f.
Eq.~(\ref{R2}). Now, $\alpha$ and the vertex next to it are measured
in the $\sigma_x$-basis. By Eq.~(\ref{R3}), $\alpha$ moves two
vertices closer to $\delta$ This procedure is repeated until
$\alpha$ and $\beta$ are merged into a single vertex $\alpha'$.
Then, in an analogous manner, $\alpha'$ is merged with $\gamma$, and
with $\beta$. Thereby, the ring of four T-junctions is converted
into a single vertex of degree 4.

{\em{Step 3}} - Creating the square lattice graph of
Fig.~\ref{perc}d. By the same method as in Step 2, the line segments
between the vertices of degree 4 are contracted. This creates a
two-dimensional cluster state on a square lattice, which is known to
be a universal resource for measurement-based quantum computation
\cite{Oneway}. Since only local measurements were used in the
reduction, the original graph state $|G\rangle$ is universal as
well.

{\em{Overhead:}} The bottleneck of the construction is to guarantee
that all junctions can be treated individually, which requires $L
\sim \log \Lambda$, c.f. Eq.~(\ref{LLamb}). As displayed in
Fig.~\ref{P1}b, a qubit in the created planar cluster state claims
an area of size $8l\times 8l$ on the original honeycomb lattice
${\cal{L}}$, and thus $\Lambda' \sim \Lambda/\log\Lambda$, as
claimed. $\Box$

Our proof thus generalizes the results in Ref.~\cite{BPerc} to
random graphs. This extends the set of cluster-type universal states
to more general 2D random graph states and beyond regular
lattices~\cite{Universal}.

\section{Monte Carlo simulations}
\label{sec:MonteCarlo}
We give the recipe for performing Monte Carlo simulations and present some results. \\
\noindent (1) First, we randomly assign every site on the honeycomb
lattice to be
either $x$, $y$ or $z$-type with equal probability. \\
\noindent (2) Second, we use the Metropolis method to sample typical
configurations. For each site we attempt to flip the type to one of
the other two with equal probability. Accept the flip with a
probability $p_{\rm accept}=\min\left\{1, 2^{|V'|-|{\cal
E}'|-|V|+|{\cal E}|}\right\}$, where $|V|$ and $|{\cal E}|$ denote
the number of domains and inter domain edges (before the modulo-2
operation on inter-domain edges; see Fig.~2 of main text),
respectively before the flip, and similarly $|V'|$ and $|{\cal E}'|$
for the flipped configuration. The counting of $|V|$ and $|{\cal
E}|$, etc. is done via a generalized Hoshen-Kopelman
algorithm~\cite{HoshenKopelman}.
For the proof of the probability ratio, see Sec.~\ref{app:Pratio}.\\

\noindent (3) After many flipping events, we measure the properties
regarding the graph structure for the domains and study their
percolation properties upon deleting edges. For the percolation, we
cut open the lattice and
investigate the percolation threshold for the typical random graphs from the Metropolis sampling.\\

\subsection{Evaluation of probability ratio}
\label{app:Pratio} In this section, we shall explain the transition
probability ratio: $p_{\rm accept}=\min\left\{1, 2^{|V'|-|{\cal
E}'|-|V|+|{\cal E}|}\right\}$, which arises from the probability for
POVM outcomes $\{a_v\}$ being $p(\{a_v\}\sim 2^{|V|-|{\cal E}|}$,
where $|V|$ and $|{\cal E}|$ denote the number of domains and
inter-domain edges (before the modulo-2 operation). The proof is
very similar to that in 1D. For convenience, we shall use spin-3/2
representation of the AKLT state. The local mapping from three
virtual qubits to one spin-3/2 is
\begin{eqnarray}
\hat{P}_{v}=\ket{1}\bra{000}+\ket{2}\bra{111}+\ket{3}\bra{W}+\ket{4}\bra{\overline{W}},
\end{eqnarray}
where we have simplified the notation for the spin-3/2 basis states:
$\ket{1}\equiv\ket{3/2,3/2}$, $\ket{2}\equiv\ket{3/2,-3/2}$,
$\ket{3}\equiv\ket{3/2,1/2}$ and $\ket{4}\equiv\ket{3/2,-1/2}$.
Moreover, $\ket{000}$, $\ket{111}$, $\ket{W}$ and
$\ket{\overline{W}}$ constitute the basis states for the symmetric
subspace of three spin-1/2 particles. The AKLT state can then be
expressed as
\begin{equation}
\ket{\psi}_{\rm AKLT}=\mathop{\bigotimes}_v \hat{P}_{v} \prod_{e=(
u,v)\in E}\ket{\phi}_{e},
\end{equation}
where $\ket{\phi}_e$ is the singlet state
$(\ket{01}-\ket{10})_{u_i,v_j}$ for the edge $e=(u,v)$ and $i, j$
specify the virtual qubit in the respective vertex.

The POVM that reduces the spin-3/2 AKLT to a spin-1/2 graph state
consists of elements $E_\mu=F_\mu^\dagger F_\mu$ such that
$\openone= E_x +E_y +E_z$, with
\begin{eqnarray}
\!\!\!\!\!\!\!\!\!\!&&\hat{F}_z=\hat{F}_z^\dagger\equiv\sqrt{\frac{2}{3}}(\ketbra{1}+\ketbra{2})=\frac{1}{\sqrt{6}}\big(S_z^2-\frac{1}{4}\big), \\
\!\!\!\!\!\!\!\!\!\!&&\hat{F}_x=\hat{F}_x^\dagger\equiv\sqrt{\frac{2}{3}}(\ketbra{a}+\ketbra{b})=\frac{1}{\sqrt{6}}\big(S_x^2-\frac{1}{4}\big),\\
\!\!\!\!\!\!\!\!\!\!&&\hat{F}_y=\hat{F}_y^\dagger\equiv\sqrt{\frac{2}{3}}(\ketbra{\alpha}+\ketbra{\beta})=\frac{1}{\sqrt{6}}\big(S_y^2-\frac{1}{4}\big),
\end{eqnarray}
where we have also expressed $\hat{F}$'s in terms of the
corresponding spin operators. The other four states other than
$\ket{1}$ and $\ket{2}$ are
\begin{eqnarray}
\!\!\!\!\!\!\!\!\!\!\!\!&&\ket{a}\equiv|S_x=3/2\rangle=\frac{1}{\sqrt{8}}(\ket{1}+\ket{2}+\sqrt{3}\ket{3}+\sqrt{3}\ket{4})\\
\!\!\!\!\!\!\!\!\!\!\!\!&&\ket{b}\equiv|S_x=\!-\!3/2\rangle=\frac{1}{\sqrt{8}}(\ket{1}-\ket{2}-\sqrt{3}\ket{3}+\sqrt{3}\ket{4})
\\
\!\!\!\!\!\!\!\!\!\!\!\!&&\ket{\alpha}\equiv|S_y=3/2\rangle=\frac{1}{\sqrt{8}}(\ket{1}-i\ket{2}+i\sqrt{3}\ket{3}-\sqrt{3}\ket{4})\\
\!\!\!\!\!\!\!\!\!\!\!\!&&\ket{\beta}\equiv|S_y=\!-\!3/2\rangle=\frac{1}{\sqrt{8}}(\ket{1}+i\ket{2}-i\sqrt{3}\ket{3}-\sqrt{3}\ket{4}).
\end{eqnarray}
They correspond to the four virtual three-spin-1/2 states (in
addition to $\ket{000}$ and $\ket{111}$) $\ket{+++}, \ket{---},
 \ket{\,\,i\,\,i\,\,i}$ and $\ket{\!-\!i,\!-\!i,\!-\!i}$.

While the outcome of POVM constructed above at each site is random
(either $x$, $y$ or $z$), outcomes at different sites may be
correlated. For a particular set of outcomes $\{a_v\}$ at sites
$\{v\}$, the resultant state is transformed to the following
un-normalized state
\begin{equation}
\ket{\psi'}=\mathop{\bigotimes}_v\hat{F}_{v,a_v} \ket{\psi}_{\rm
AKLT},
\end{equation}
with the probability being
\begin{equation}
p_{\{a_v\}}=\ipr{\psi'}{\psi'}/\ipr{\psi}{\psi}_{\rm AKLT}.
\end{equation}
As $\hat{F}$'s are proportional to projectors, in evaluating the
relative probability for two sets of outcome ${\{a_v\}}$ and
${\{b_v\}}$, one has
\begin{equation}
p_{\{a_v\}}/p_{\{b_v\}}=\bra{\psi}\mathop{\bigotimes}_v\hat{F}_{v,a_v}
\ket{\psi}_{\rm
AKLT}\big/\bra{\psi}\mathop{\bigotimes}_v\hat{F}_{v,b_v}
\ket{\psi}_{\rm AKLT}, \label{eqn:pratio}
\end{equation}
where we have used $\hat{F}_{v,a}^2 \sim \hat{F}_{v,a}$. In order to
evaluate the probability ratio for two different sets of
configuration, we first note that
\begin{eqnarray}
&&\hat{F}_x \hat{P}\sim \ket{a}\bra{+++}+\ket{b}\bra{---}\\
&&\hat{F}_y \hat{P}\sim \ket{\alpha}\bra{i\,i\,i}+\ket{\beta}\bra{-i-i-i}\\
&&\hat{F}_z \hat{P}\sim \ket{1}\bra{000}+\ket{2}\bra{111}.
\end{eqnarray}
The spin-3/2 state is transformed by $\bigotimes_v F_{v,a_v}$ to an
effective spin-1/2 one, with the two levels being labeled by
$(a,b)$, $(\alpha,\beta)$, or $(1,2)$, depending on which $\hat{F}$
is applied. The probability $p_{\{a_v\}}$ is essentially obtained by
summing the square norm of the coefficients for all possible
spin-1/2 constituent basis states (e.g. $\ket{a\,b+0\,i\dots}$ is a
basis state). First we need to know how many different constituent
states, and the number is related to how many effective spin-1/2
particles we have. For the sites that have same type of outcome
($x$, $y$ or $z$), they basically form a superposition of two
Ne\'el-like states, thereby corresponding to an effective spin-1/2
particle. This can be seen from the valence-bond picture that, e.g.,
for $r,s\in \{0,1\}$ we have
$\bra{rs}01-10\rangle=\pm\delta_{r,1-s}$. On the other hand, for
$r\in\{0,1\}$ and $s\in\{+,-\}$, $\bra{rs}01-10\rangle=\pm
1/\sqrt{2}$, which is $1/\sqrt{2}$ smaller than if $r$ and $s$ are
the indices in the same basis. This means that all four combinations
$\{0+,0-,1+,1-\}$ occur with equal amplitude up to a phase. (Similar
consideration applies to other combinations of bases.) Therefore,
the number of effective spin-1/2 particles is given by the number of
domains, which we label by $|V|$. Notice that we have assumed that
any domain does not contain a cycle with odd number of original
sites, as no Ne\'el state can be supported on such a cycle (or
loop). Configurations with domains that contain a cycle with odd
number need to be removed. Fortunately, as the honeycomb lattice is
bi-partite, therefore any cycle must contain even number of sites
and we do not to deal with the above complication.

 What about the amplitude for each spin configuration? Furthermore,
what is the probability of obtaining a particular set of outcome
$\{a_v\}$? We have seen that for each inter-domain edge there is a
contribution to a factor of $1/\sqrt{2}$ in the amplitude (as the
end sites of the edge correspond to different types). Thus, the
amplitude for each spin configuration gives an overall value
(omitting the phase factor) of $2^{-|{\cal E}|/2}$ and hence a
probability weight $2^{-|{\cal E}|}$, where $|{\cal E}|$ counts the
number of inter-domain edges. As there are $2^{|V|}$ such
configurations, we have the norm square of the resultant spin-1/2
state being proportional to $p\sim 2^{|V|-|{\cal E}|}$. For
convenience, we have assume the lattice is periodic, but the
argument holds for  open boundary condition in which the spin-3/2's
at the boundary are either (1) suitably linked by to one another,
preserving the trivalence or (2) terminated by spin-1/2's. In the
Appendices, we have provided an alternative derivation of the
probability expression.
\begin{figure}
\hspace{-0.5cm}{\includegraphics[width=9cm]{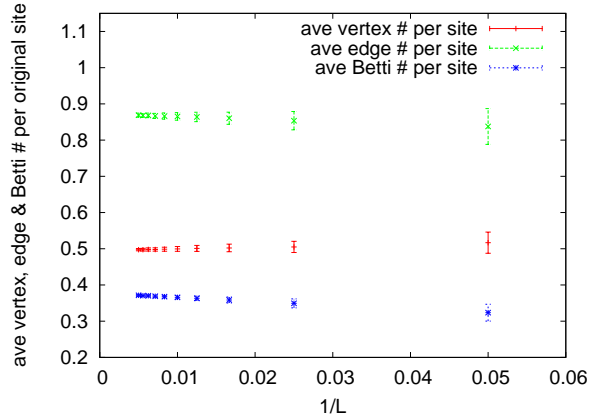}}
        \caption{(color online)  Average vertex (or domain) number, average edge number, and average Betti number in  the typical random graphs original lattice site vs. $L$.
       The total number of sites is $N=L^2$.  This
shows the number of domains, the number of interdomain Ising
interaction, and the number of independent loops in the resultant
graph all scale with the system size of the original honeycomb
lattice.} \label{fig:ave}
\end{figure}
\subsection{Discussions of simulation results}

\begin{figure}
\hspace{-0.5cm}{\includegraphics[width=9cm]{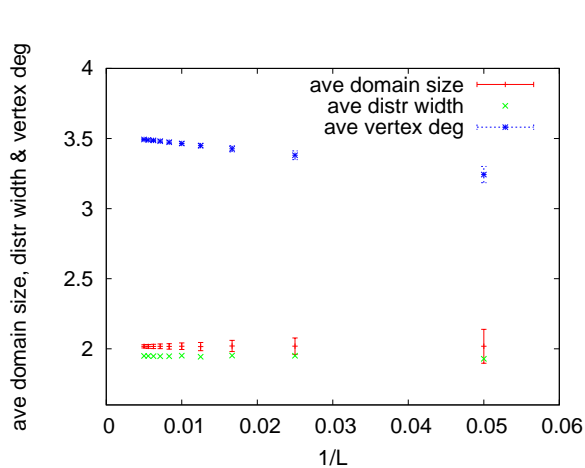}}
 \caption{(color online)  Average domain size (i.e., number of original sites in a domain), average width of domain size distribution, and average degree of a vertex in the typical random graphs vs. $L$, where $L^2$ is the total number
 of sites in the honeycomb lattice. For better discernibility of the two lower sets of data, we suppress
 the errorbars for one of them. This set of data was first shown in Ref.~\cite{WeiAffleckRaussendorf11},
 and we have reproduced it here for the sake of completeness.} \label{fig:avdeg}
\end{figure}

We have analyzed lattices of size up to $200 \times 200$ sites. As
shown in Fig.~\ref{fig:ave}, the size dependence of average vertex
number, average edge number, and average Betti number
$B$~\cite{Betti} of the random graphs formed by domains relative to
the original lattice size behaves as follows: $|\bar{V}|= 0.495(2)
L^2$, $|\bar{E}|=0.872(4) L^2$, and $\bar{B}= 0.377(2) L^2$, where
$L$ is related to the total number of sites in the original
honeycomb lattice $N=L\times L$. This shows that the typical random
graph of the graph state retains macroscopic number of vertices,
edges, and cycles, giving strong evidence that the state is a
universal resource. Figure~\ref{fig:avdeg} shows the average degree
of a vertex vs. inverse system length $1/L$ for the random graphs,
as well as the average numbers of the original sites contained in a
typical domain. The average vertex degree extrapolates to
$\bar{d}\approx3.52(1)$ for the infinite system. This compares to
$4$ for the square lattice and $3$ for the honeycomb lattice.

\begin{figure}
\begin{tabular}{c}
(a)\\\hspace{-0.5cm}{\includegraphics[width=9cm]{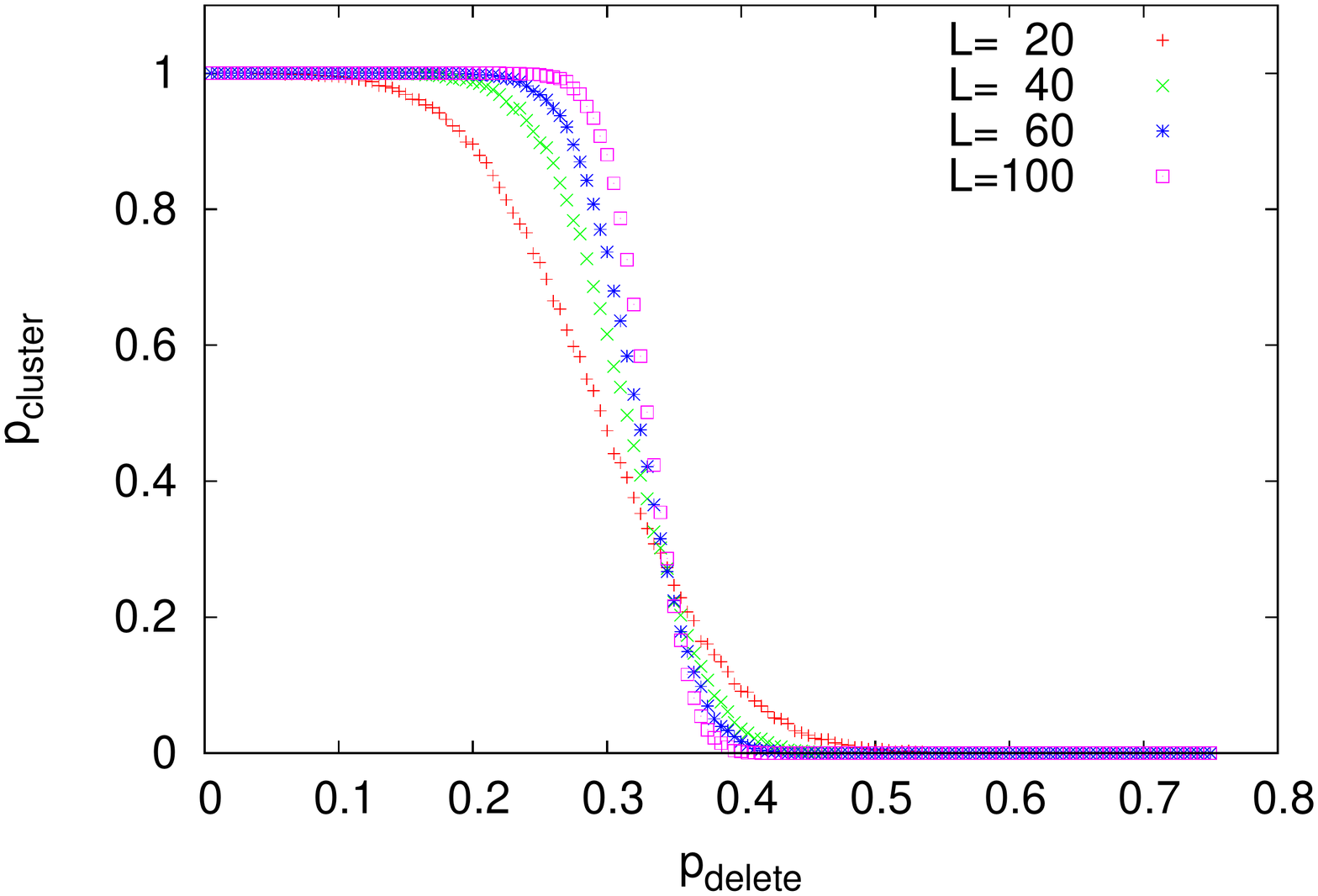}}\\
(b)\\ \hspace{-0.5cm}{\includegraphics[width=9cm]{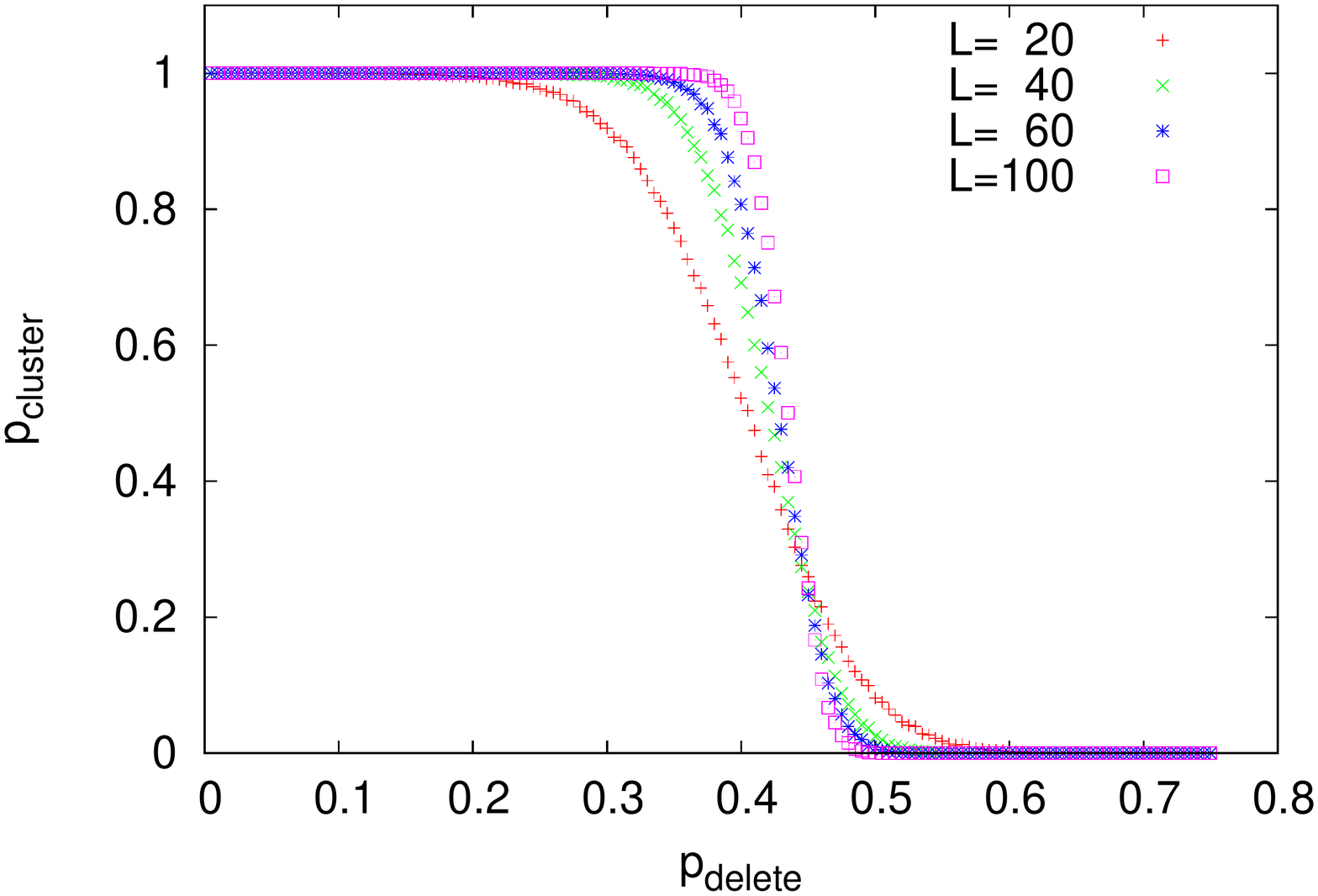}}
\end{tabular}
        \caption{(color online)  Percolation study of the graph formed by the domains: probability of
        a spanning cluster $p_{\rm cluster}$ vs. the probability to delete a vertex (top panel a) or an edge (bottom panel b) $p_{\rm delete}$.
        The threshold
        for destroying the spanning cluster is around $p_{\rm delete}\approx 0.33$ in deleting vertices
        and $p_{\rm delete}\approx 0.43$ in deleting edges. This
shows that the graph without deleting any vertex or edge is deep in
the percolated (i.e., connected) phase. We note that the result of
site percolation is reproduced from
Ref.~\cite{WeiAffleckRaussendorf11}, and the additional bond
percolation result presented here is consistent with the picture
that the typical random graphs lie somewhat between the honeycomb
and the square lattice. } \label{fig:per}
\end{figure}

In order to show the stability of the random graph, we investigate
how robust it is upon, e.g., deleting vertices (or edges)
probabilistically, i.e.,  performing the site (or bond) percolation
simulations. As shown in Fig.~\ref{fig:per}a, it requires the
probability of deleting vertices to be as high $p_{\rm delete}=
0.33(1)$ (i.e., percolation threshold $p_{\rm c}= 0.67(1)$) in order
to destroy the spanning property of the graph. This lies between the
site percolation thresholds $\approx 0.592$ of the square lattice
and $\approx 0.697$ of the honeycomb lattice. For bond percolation
as shown in Fig.~\ref{fig:per}b, it takes a probability of $p_{\rm
delete}= 0.43(1)$ (i.e., percolation threshold $p_{\rm c}= 0.57(1)$)
to destroy the spanning property of the graph. Again, this threshold
lies between that of the square lattice ($1/2$) and that of the
honeycomb lattice ($\approx 0.652$). This shows that there exists
many paths (proportional to the system's linear size) on the random
graphs that can be used to simulate one-qubit unitary gates on as
many logical qubits and entangling operations among them. We remark
that percolation argument was previously employed by Kieling,
Rudolph, and Eisert in establishing the universality of using
nondeterministic gates to construct a universal cluster
state~\cite{KielingRudolphEisert}.

\begin{figure}
\vspace{-0.5cm} {\includegraphics[width=8.5cm]{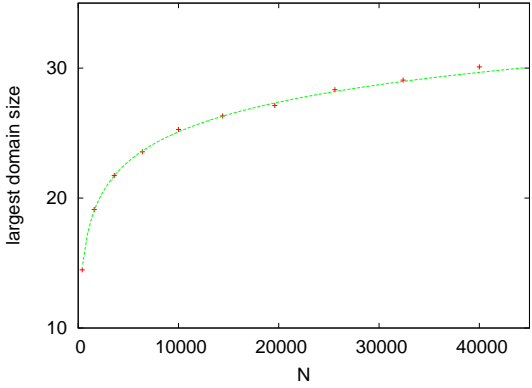}}
\vspace{-1cm}
 \caption{(color online)  The largest domain size in the typical graphs
 vs. $L$, with $N=L^2$ being the total number
 of sites. The fitted curve to the largest domain size is
 $3.337\ln(N)-5.566$. The result is reproduced from Ref.~\cite{WeiAffleckRaussendorf11}.} \label{fig:domainsize}
\end{figure}

Let us also examine the two conditions listed in
Sec.~\ref{2Dreduc}.\\
 \indent {\em{Condition C1.}} For all POVM outcomes sampled from, the size of the largest
 domain was never macroscopic and it can at best be logarithmic in the original system size; see Fig.~\ref{fig:domainsize}.
 The average number
 of sites $v \in V({\cal{L}})$ contained in  a typical domain, when extrapolated to
 the infinite system, is $2.02(1)$; see Fig.~\ref{fig:avdeg}. Our numerical simulations thus show that
 condition C1 holds.

{\em{Condition C2.}} For all of the POVM outcomes sampled, a
horizontal and a vertical traversing path through the resulting
graphs $G({\cal{A}})$ always existed (without deleting any vertex or
edge). Our numerical simulations show that our random graph are deep
in the supercritical phase and thus condition C2 holds.

In addition, a necessary condition for the computational universality of the
graph states $|G({\cal{A}})\rangle$ is that typical graphs $G({\cal{A}})$ are
not close to trees, because MBQC on tree-like graphs can be efficiently
classically simulated \cite{Shi}. For typical graphs $G({\cal{A}})$, we find
that the Betti number (which is zero for trees) is proportional to the size of
the initial honeycomb graph, with $\bar{B}= 0.377(2) N$.

{\em{Robustness.}} We now quantify how deep typical graphs
$G({\cal{A}})$ are in the connected phase of the percolation
transition. A first measure is the average vertex degree. A
heuristic argument based on a branching process suggests that a
graph has a macroscopic connected component whenever the average
vertex degree is $\bar{d}>2$. This criterion is exact for random
graphs of uniform degree~\cite{Perc}. It also holds surprisingly
well for lattice graphs~\cite{footnote}, which are the least random.
In our case, the typical graphs $G({\cal{A}})$ have an average
degree of 3.52, suggesting that the system is deep in the connected
phase, which is confirmed by the percolation simulations. The
existence of finite percolation thresholds (for both site and bond
percolation) discussed earlier further supports the robustness of
the connectedness.

\section{Concluding remarks}
\label{sec:Conclude}
 We investigated the measurement-based quantum
computation on the AKLT states. First we provided an alternative
proof that the 1D spin-1 AKLT state can be used to simulate
arbitrary one-qubit unitary gates. We extended the same formalism
and demonstrated that the spin-3/2 AKLT state on a two-dimensional
honeycomb lattice is a universal resource for measurement-based
quantum computation by showing that a 2D cluster state can be
distilled by local operations. Along the way, we connected the
quantum computational universality of 2D random graph states to
their percolation property and showed those 2D graph states whose
graphs are in the supercritical phase are indeed universal resources
for MBQC.

The key ingredient that has enabled our proof of computational
universality for the (spin-3/2) 2D AKLT state on the honeycomb
lattice is the generalized measurement in Eq.~(\ref{POVM2}). How
about the case of (spin-2) 2D AKLT state on the square lattice or
any other lattices beyond trivalence is universal for MQBC.
  The AKLT spin-2 particle
can be regarded as four virtual qubits in the symmetric subspace.
Hence a naive extension of the POVM prompts us to consider the
follow operators:
\begin{subequations}
  \begin{eqnarray}
{F}_{z}&=&(\ketbra{0^{\otimes 4}}+\ketbra{1^{\otimes 4}}) \\
{F}_{x}&=&(\ketbra{+^{\otimes 4}}+\ketbra{-^{\otimes 4}})\\
{F}_{y}&=&(\ketbra{i^{\otimes 4}}+\ketbra{-\!i^{\otimes 4}}).
\end{eqnarray}
\end{subequations}
Unfortunately, $\sum_\alpha F^\dagger F^\dagger$ is not proportional
to the projection onto the symmetric subspace. However, we can
 consider additionally the four states $\ket{\gamma_k}$ ($k=1,..,4$)
such that their Bloch vectors point in the four diagonal directions
of a cube, i.e., $(1,1,1)/\sqrt{3}$, $(-1,1,1)/\sqrt{3}$,
$(-1,-1,1)/\sqrt{3}$, and $(1,-1,1)/\sqrt{3}$, respectively.
Together with their corresponding conjugate states
$\ket{\bar{\gamma}_k}$ having opposite vectors, we have four other
sets of projections:
\begin{equation}
G_{k} \equiv \ketbra{\gamma_k^{\otimes 4}}
+\ketbra{\bar{\gamma}_k^{\otimes 4}}.
\end{equation}
It can be checked that
\begin{equation}\frac{8}{24} \sum_{\alpha=x,y,z}F^\dagger_\alpha F_\alpha +
\frac{9}{24}\sum_{k=1}^4G^\dagger_{k}G_{k}=P_S,\end{equation}
 where $P_S$ is the projection operator onto the symmetric subspace of four qubits~\cite{LiEtAl}.
However, such a generalized measurement would yield four additional
pairs of states $\{\gamma_k,\bar{\gamma}_k\}$ which are not mutually
unbiased to one another nor to the eigenstates of the three Pauli
operators. Due to this complication, whether the 2D AKLT state on
the {\it square\/} lattice is universal for MQBC remains open.

\medskip
\noindent {\bf Acknowledgment.} This work was supported by NSERC,
CIFAR, the Sloan Foundation, and the C.N. Yang Institute for
Theoretical Physics.

\appendix
\section{Calculation of probability of a particular POVM outcome using Arovas-Auerbach-Haldane techniques}
In this appendix we provide an alternative formulation to the
calculation of POVM outcome probability. This formalism has the
potential of being applicable to a more general case. We give only
the important ingredients here.

 Arovas,
Auerbach and Haldane (AAH)~\cite{Arovas} show how to represent
arbitrary AKLT states as Boltzmann weights for nearest neighbour
statistical mechanical models in the same spatial dimension as the
quantum problem and how to represent calculations of equal time
ground state expectation values classically. We are interested in
two cases: the $S=1$ one-dimensional case and the $S=3/2$ honeycomb
lattice case.

 In both cases the operators of interest are proportional to projection operators
onto maximal $|S^z|$:
\bea F_\nu&\equiv& (S_\nu)^2/\sqrt{2},\ \ (S=1)\\
&\equiv& [(S_\nu)^2-1/4]/\sqrt{6},\ \  (S=3/2),\eea where $\nu=x,y\,{\rm
or}\,z$ and, for convenience, we have rescaled the prefactor in the
definition of $F$'s.
For general spin S the operators $S^a$ are represented first
in terms of Schwinger bosons, $a$, $a^\dagger$, $b$, $b^\dagger$,
then in terms of co-ordinates and derivatives $u$, $v$, $\partial_u$,
$\partial_v$ acting on homogeneous polynomials of $O(2S)$. The operator
$(S^z)^2$ is:
\bea && (S^z)^2=(1/4)(a^\dagger a-b^\dagger b)^2=(1/4)(\partial_uu-\partial_vv)^2\nonumber \\
&&=(1/4)(\partial_u^2u^2+\partial_vv^2-2\partial_u\partial_vuv-\partial_uu-\partial_vv)
.\eea We now use the prescription of Arovas, Auerbach and Haldane:
\bea && \langle\psi '|\partial_u^k\partial_v^lu^{k+j}v^{l-j}|\psi
\rangle \nonumber \\ &&=\left[\prod_{m=2}^{k+l+1}(2S+m)\right]
\langle\psi '|u^{*k}v^{*l}u^{k+j}v^{l-j}|\psi \rangle\label{id0}
\eea for any states $|\psi '\rangle$ and $|\psi \rangle$ in the spin
$S$ Hilbert space.

To prove Eq.~(\ref{id0}), note that a complete set of states for the
spin $S$ Hilbert space is given by $u^{S+m}v^{S-m}$ which are
eigenstates of $S^z$ with eigenvalue $m=-S, -S+1, \ldots S$. To
prove Eq. (\ref{id0}) for $j=0$, we wish to prove: \bea
\!\!\!\!\!\!\!\!\!\!&&\int d^2\Omega
u^{*S+m}v^{*S-m}\partial_u^k\partial_v^lu^{k+S+m}v^{l+S-m} \nonumber
\\ \!\!\!\!\!\!\!\!\!\!&=&  \left[ \prod_{r=2}^{k+l+1}(2S+r)\right] \int d^2\Omega
|u|^{2(S+m+k)}|v|^{2(S-m+l)}.\label{id}\eea To prove this, we use
the identity: \bea I_{p,q}&\equiv& \int d^2\Omega
|u|^{2p}|v|^{2q} \nonumber\\
&=&2\pi (1/2)^{p+q}\int_{-1}^1dx(1+x)^p(1-x)^q \nonumber \\
&=&4\pi p!q!/(p+q+1)!\eea
Thus the left hand side of Eq. (\ref{id}) may be written:
\bea LHS&=&{(S+m+k)!\over (S+m)!}{(S-m+l)!\over (S-m)!}I_{S+m,S-m}\nonumber \\
&=&4\pi {(S+m+k)!(S-m+l)!\over (2S+1)!} \nonumber \\
&=&\left[ \prod_{r=2}^{k+l+1}(2S+r)\right]I_{S+m+k,S-m+l}\label{id2}
\eea
which is the RHS.
Furthermore, all off-diagonal matrix elements vanish for both the left and right hand side of the identity
 in Eq.~(\ref{id0}) for $j=0$. That follows
since $\psi^*_m\psi_{m'}\propto e^{i(m'-m)\phi}$ where $\phi$ is the
azimuthal angle for the integration over the sphere. Neither
inserting the operator on the left hand side of Eq.~(\ref{id0}) nor
multiplying by the function on the right hand side changes this
azimuthal angle dependence, implying vanishing integrals. While it
may appear that this proof only holds for $j=0$ in Eq.~(\ref{id0})
it actually covers the case of general $j$. In general,
Eq.~(\ref{id}) gives the $\langle \psi_m|\ldots |\psi_{m-j}\rangle$
matrix elements of the identity in Eq.~(\ref{id0}), which are the
only non-zero matrix elements. Furthermore the identity immediately
generalizes to an arbitrary product on different lattice sites: \bea
&&\langle\psi
'|\prod_i\partial_{u_i}^{k_i}\partial_{v_i}^{l_i}u_i^{k_i+j_i}v_i^{l_i-j_i}|\psi
\rangle =\left[\prod_i\prod_{m=2}^{k_i+l_i+1}(2S+m)\right]\nonumber \\
&& \times \langle\psi
'|\prod_i|u_i^{*k_i}v_i^{*l_i}u_i^{k_i+j_i}v_i^{l_i-j_i}|\psi
\rangle \eea since we may simply extend the above argument to the
basis states $\prod_iu_i^{S+m_i}v_i^{S-m_i}$ for which the matrix
elements simply factorize. Since we have proved this identity for a
complete set of states  it follows for any states $|\psi \rangle$,
$|\psi '\rangle$ in the spin-$S$ Hilbert space, including the AKLT
states.

Eq.~(\ref{id0}) gives: \bea
(S^z)^2&=&(1/4)(2S+2)(2S+3)[|u|^4+|v|^4-2|u|^2|v|^2]\nonumber \\ &&
-(1/4)(2S+2)(|u|^2+|v|^2)\nonumber \\
&=&(1/4)(2S+2)(2S+3)[|u|^2-|v|^2]^2\nonumber \\ &&
-(1/2)(S+1)(|u|^2+|v|^2).
\eea
Using $u=\cos (\theta /2) e^{i\phi /2}$, $v=\sin (\theta /2) e^{-i\phi /2}$,
where $\theta$ and $\phi$ are the polar and azimuthal angle on the
unit sphere, this becomes:
\be (S^z)^2=(1/4)(2S+2)(2S+3)(\Omega^z)^2-(1/2)(S+1)\ee
where $\Omega_z=\cos \theta$ is the projection of the unit vector onto the $z$-axis.
A simple explicit calculation similar to this one shows that, for
$\nu=x$, $y$ or $z$: \be
(S_\nu)^2=(1/4)(2S+2)(2S+3)(\Omega^\nu)^2-(1/2)(S+1)\ee as expected
by SO(3) symmetry. This is a somewhat surprising formula in that the
classical quantities are not positive semi-definite. Note that this
formula is valid independent of the wave-function. The projection
operators thus become:
\bea F_\nu&=&[5(\Omega^\nu)^2-1]/\sqrt{2},\ \  (S=1)\\
&=&\sqrt{3/8}\,[5(\Omega^\nu)^2-1],\ \  (S=3/2).\eea Remarkably the
projection operators are the same for $S=1$ and 3/2 up to an
unimportant normalization factor.

The AKLT state can be written, in Schwinger boson notation as: \be
|\psi \rangle_{\rm AKLT}=\prod_{\langle
i,j\rangle}(a^\dagger_ib^\dagger_j-a^\dagger_jb^\dagger_i)|{\rm
vacuum}\rangle \ee
corresponding to
\be \psi (u_i,v_i)=\prod_{\langle i,j\rangle}(u_iv_j-v_iu_j)\ee
where the product is over all pairs of neighboring
sites $(i,j)$ . The square of the wave-function is:
\be |\psi (u_i,v_i)|^2\propto \prod_{\langle i,j\rangle}[1-\hat \Omega_i\cdot \hat \Omega_j].\ee
Actually, we need to be more precise about boundary
conditions here. These details will be discussed below. Using
the form of the AKLT state we wish to calculate: \be p_{a_1a_2\ldots
a_N}\equiv {\cal N}_S{1\over Z}\prod_{i=1}^n\int d\hat \Omega_i
[5(\Omega^{a_i}_i)^2-1]\prod_{\langle
j,k\rangle}[1-\hat\Omega_j\cdot \hat\Omega_j]\ee where $Z$ is the
same integral without the $[5(\Omega^{a_i}_i)^2-1]$ factors and
${\cal N}_S=(1/2)^n$ for $S=1$ and $(1/3)^n$ for $S=3/2$. Note that
the inserted operators are $F_\nu/\sqrt{2}$ for the $S=1$ case
and $\sqrt{2/3}F_\nu$  for the $S=3/2$ case, normalized so that the
sum over $\nu$ gives the identity operator, ensuring the proper normalization
of the probability distribution.
In both cases we
can evaluate this by multiplying out $\prod_{\langle
j,k\rangle}[1-\hat\Omega_j\cdot \hat\Omega_k]$.

Carrying out this we arrive at the same conclusion of the
probability expressions for 1D chain and 2D honeycomb cases as
before, as we show below.

\begin{figure}
 \includegraphics[width=8cm]{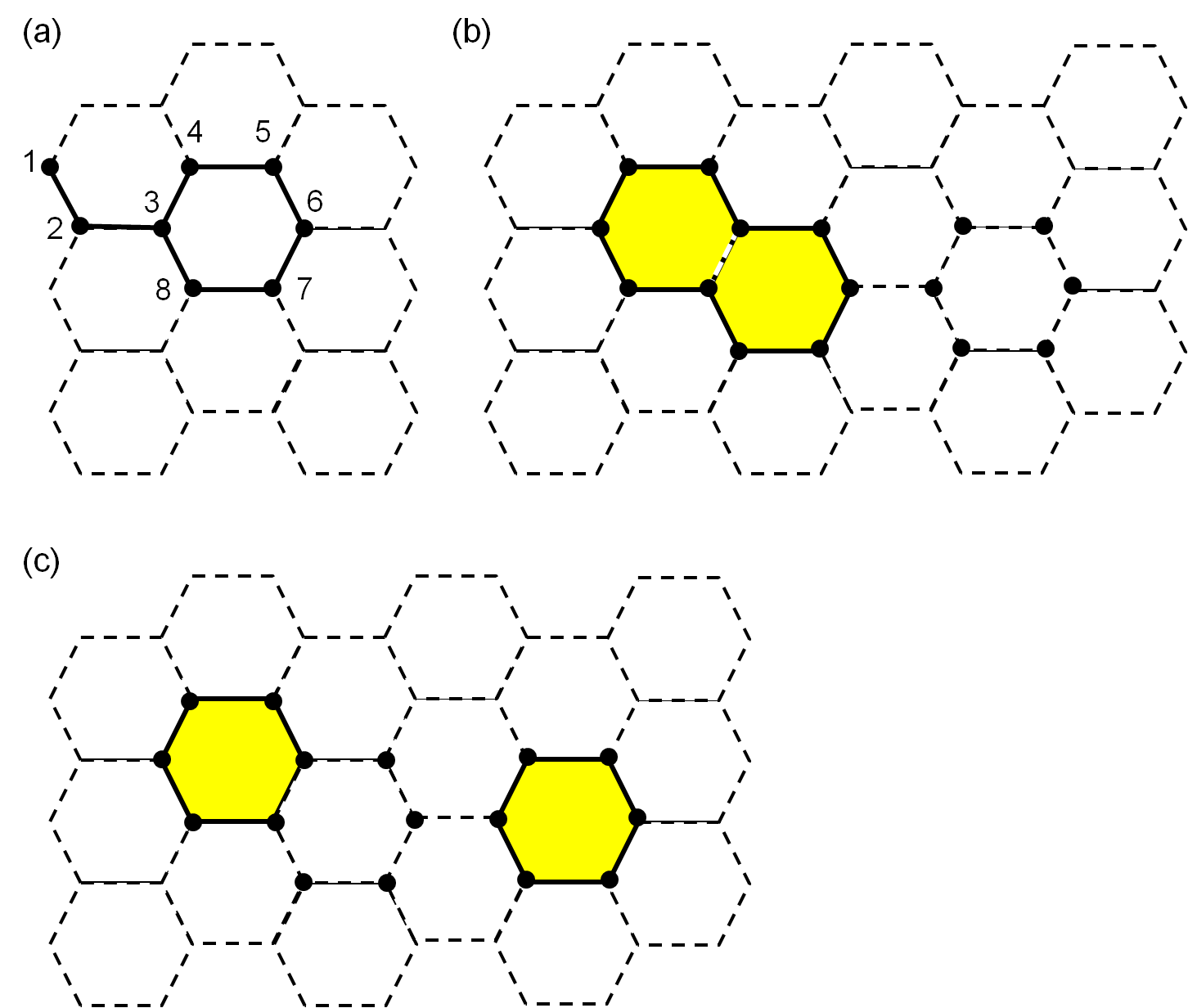}
  \caption{\label{fig:loopsOB}
Illustration of loops. (a) A domain consisting of 8 vertices, 8
edges and 1 face. The numbers indicate one possible order in which
the domain could be grown. When the first vertex is added, $V=1$,
$E=F=0$. As vertices 2 to 7 are added both $V$ and $E$ increase by
1. When the 8$^{\hbox{th}}$ vertex is added, $V$ and $F$ increase by
1 and $E$ increases by 2. (b) In this domain, which contains $F=3$
faces, two of them are selected, as indicated by shading. The
corresponding set of loops is a single loop surrounding the two
adjacent faces as indicated by heavy lines. This domain contains
$V=16$ vertices and $E=18$ edges, obeying $F=E-V+1$. (c) This is the
same domain as in the previous figure (b) but a different subset of
faces is selected. Now the set of loops consists of 2 loops as
indicated by heavy lines. }
\end{figure}
\section{S=1, 1 dimension}
We first consider the 1D S=1 case as a warm-up.
\subsection{Open Boundary Conditions}
Consider a chain of $n$ spin-1's on sites $i=1,2,\ldots n$ with 2
additional S=1/2's at sites 0 and n+1 to remove the ``dangling
bonds''. Then the AKLT ground state is: \be |\psi
\rangle_0=\prod_{i=0}^n(a^\dagger_ib^\dagger_{i+1}-a^\dagger_{i+1}b^\dagger_i)|0\rangle.\ee
Thus: \be |\psi_0|^2=\prod_{i=0}^n[1-\hat \Omega_i\cdot
\hat \Omega_{i+1}].\ee We only make projective measurements on the sites
containing spin-1's, at  $1,2,\ldots n$. In this case we may replace
each factor $1-\hat \Omega_i\cdot \hat \Omega_j$ by $1$ because all other terms
in the expansion contain a single power of one or more $\hat
\Omega_i$ vector and thus give zero after integrating over $\hat
\Omega_i$. Using:
\be \langle(\Omega^a)^2\rangle=(1/3)\langle(\hat \Omega )^2\rangle=1/3,\  \
 \langle(5\Omega^a)^2-1\rangle=2/3,
\ee
we obtain a constant: \be  P^{a_1a_2\ldots
a_n}=(1 /3)^n\ee
independent of the $a_i$'s.
\subsection{Periodic Boundary Conditions}
Now we consider $n$ sites, all with spin-1's and couple site $n$ to
site $1$. This is a useful warm up for the 2D case because there is
now one closed loop, i.e., one other term in the expansion can give
a non-zero integral: $(-1)^n\prod_{i=1}^n\Omega_i\cdot
\Omega_{i+1}$. Now we need the integral: \be \int d\hat \Omega
[5(\Omega^a)^2-1]\Omega^b\Omega^c.\ee Clearly this vanishes unless
$b=c$. Note that: \be \langle(\Omega^z)^4\rangle=\langle\cos^4\theta
\rangle=(1/2)\int_{-1}^1dx x^4=1/5\ee and \bea
\langle(\Omega^x)^2(\Omega^y)^2\rangle&=&\int_0^{2\pi}{d\phi\over
2\pi}\cos^2\phi \sin^2\phi \int_{-1}^1{dx\over
2}(1-x^2)^2\nonumber\\
&=&1/15.\eea Thus we obtain the remarkable identity: \be
\langle[5(\Omega^a)^2-1]\Omega^b\Omega^c\rangle={2\over
3}\delta^{bc}\delta^{ab}.\ee The integral vanishes unless $a=b=c$ in
which case it has the same value as $\langle(5\Omega^a)^2-1\rangle$.
The product $(-1)^n\prod_{i=1}^n\hat \Omega_i\cdot\hat \Omega_{i+1}$
contains sums over $n$ indices.  However, for the integrals to be
non-zero all $a_i$ indices  must equal each other. In this case the
multiple integral has exactly the same value as when the product is
not present, giving: \be  P^{a_1a_2\ldots a_n}=(1/3)^n{[1+(-1)^n
\delta_{a_1a_2}\delta_{a_2a_3}\ldots \delta_{a_na_1}]\over
1+(-1)^n(1/3)^{n-1}}.\ee Here we have used the fact that the
partition function also obtains a contribution from
$(-1)^n\prod_{i=1}^n\Omega_i\cdot \Omega_{i+1}$ giving the second
term in the denominator and the reason that it is three times larger
than $1/3^n$ is because of three possibilities $a=x,y,z$, or more
precisely, \begin{equation}
 (1/4\pi )^n\int\prod_{i=1}^n d^2\Omega_i\,
\vec\Omega_i \cdot \vec \Omega_{i+1}=(1/3)^{n-1}, \end{equation}
where $\vec \Omega_{n+1}\equiv \vec \Omega_1$. Thus,
$P^{a_1a_2\ldots a_n}$ is nearly constant again except that in the
one case where $a_1=a_2=\ldots a_n$ it is twice as big if $n$ is
even or zero if $n$ is odd. This result agrees precisely with that
obtained by other methods in sub-section Sec.~\ref{sec:1Dprob}.

\begin{figure}[ht!]
 \includegraphics[width=8cm]{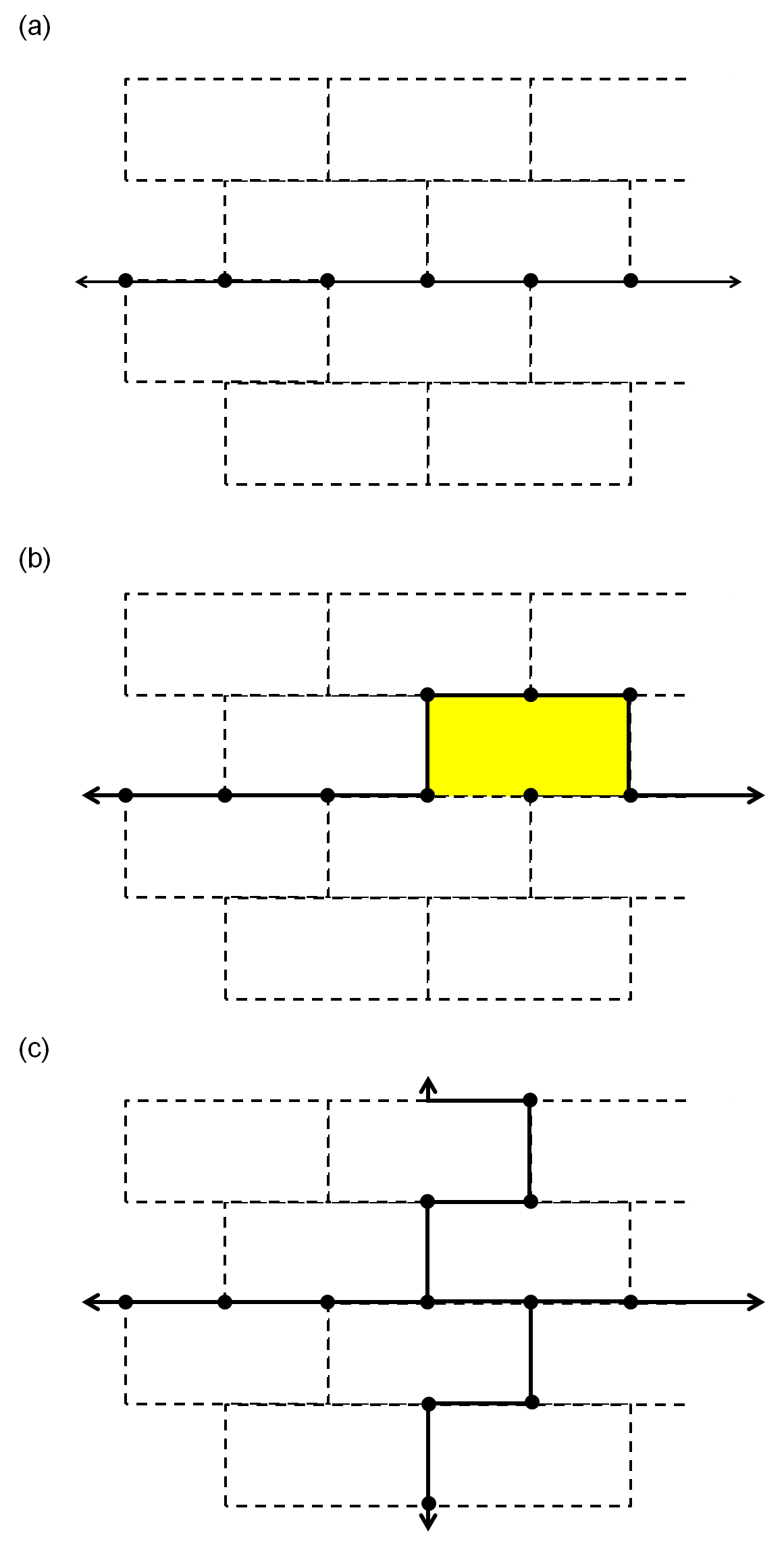}
  \caption{\label{fig:loopsPB}
Illustration of loops in the case of the periodic boundary
condition. A honeycomb lattice with periodic boundary conditions,
drawn as a ``brick wall'' lattice for convenience. (a) A topological
loop is shown. The arrows indicate an edge in the domain between
vertices at the left and right hand sides of the lattice. (b) The
set of loops (one loop in this case) is shown for a domain
containing one topological loop and one face sharing on edge with
the topological loop, corresponding to the case in which the face
and the topological loop are chosen.  (c)A domain containing two
topological loops, with $W=2$. It can be seen that there are 3
possible loops, corresponding to 4 sets of loops, and zero faces. }
\end{figure}

\section{S=3/2, 2 dimensions}
\subsection{Open Boundary Conditions}
Consider an arbitrary finite segment of a honeycomb lattice,
consisting  of $n$ spins; it could have zig-zag and armchair edges
or disordered ones, for example. Spins on the boundary will
generally be coupled to either 2 or 3 other spins - 2 for a zig-zag
edge and 3 for an armchair edge, for example. In all cases where a
boundary spin is only coupled to 2 other spins, couple it to a
boundary S=1/2 spin.  Let the total number of spins, including the
S=1/2 spins on the boundary be $M$. Then the square of the AKLT
ground state is: \be |\psi_0|^2=\prod_{\langle i,j\rangle}[1-\hat
\Omega_i \cdot \hat \Omega_j].\ee The product is over all nearest
neighbours, as usual including both S=3/2 and S=1/2 spins. We only
do the POVM on the S=3/2 spins. Since each of the boundary S=1/2
spins couples to only one other (S=3/2) spin,  we may replace
$[1-\hat \Omega_i\cdot \hat \Omega_j]$ by $1$ for each factor
involving an S=1/2 spin in calculating $P^{a_1a_2\ldots a_n}$.
Following the above reasoning, when we take the $1$ term in the
expansion of $\prod_{\langle i,j\rangle}[1-\hat \Omega_i \cdot \hat
\Omega_j]$, we get: \be P^{a_1a_2\ldots a_n}={1\over Z}(4\pi
)^M(1/3)^N(2/3)^N+\ldots \ee There will be many additional terms in
this case, unlike the D=1 case. Each additional term must correspond
to a set of closed loops on the lattice, with zero or two lines
entering each of the S=3/2 sites. These loops never involve the
S=1/2 boundary sites. These loops can never cross each other but we
can have loops inside loops. Such a contribution only exists when
all the $a_i$'s for sites on a given loop have the same value. Each
such term  makes an equal contribution to $P^{a_1a_2\ldots a_n}.$
Thus we simply need to calculate the number of sets of closed loops
with equal $a_i$'s for a given configuration $a_1,a_2,\ldots a_n$.

To do this it is convenient to divide up all sites on the lattice
into domains such that $a_i$ has the same value for all sites in a
domain and all sites in a domain are the nearest neighbor of at
least one other site in the domain. (Here sites refers to the sites
with S=3/2 spins only.)  The number of sites in a domain can range
from 1 to $n$, in principle, although we expect that typical domains
are microscopic.  We draw a line between all nearest neighbors in
each domain.  We may identify a unique number of faces with each
domain, $F_i$ and a total number of faces, $F=\sum_iF_i$ with a
given configuration.  A face, inside a domain, is an
elementary hexagon which is completely surrounded by
$6$ edges belonging to that domain. Thus, it is impossible
to move from the interior of a face to its exterior (either inside
the domain or not) without crossing an edge belonging to its domain.
The total number of sets of closed loops, $N_L$  is then simply \be
N_L=2^F.\label{N_L}\ee
This follows because there is a unique set of loops which surrounds
any subset of the faces.
See Fig.~\ref{fig:loopsOB}.

The $i^{\hbox{th}}$ domain will also have a number of edges, $E_i$
and a number of vertices, $V_i$.  It can be seen that: \be
F_i=E_i-V_i+1.\label{FEV}\ee This can be seen by induction, growing
the domain site by site, always adding new sites which are nearest
neighbors of at least one previous site. After drawing the first
site, $V_i=1$ and $E_i=F_i=0$, so Eq. (\ref{FEV}) is obeyed.  When
we add the next site, we increase both $E_i$ and $V_i$ by 1, without
changing $F_i$. This goes on for a while but we may eventually add a
site which is the nearest neighbor of 2 previous sites. At that
step, $V_i$ increases by 1, $E_i$ increases by 2 and $F_i$ increases
by 1 since we are then closing a loop, making a new face. Thus
Eq.~(\ref{FEV}) remains true at each step as we grow the domain,
completing the proof.

Suppose we define a new, random lattice, by collapsing each domain
down to one vertex, with an arbitrary number of edges, inherited
from the original lattice, connecting the various domains. Let $V'$
by the number of vertices of this random lattice and $E'$ be the
number of edges.  (We ignore the S=1/2 boundary spins here.) Then:
\be V'=n-\sum_i(V_i-1)\ee since $V_i$ sites are reduced to 1 at the
$i^{\hbox{th}}$ domain. Similarly if $n_E$ is the total number of
edges connecting S=3/2 spins in the original honeycomb lattice, then
\be E'=n_E-\sum E_i.\ee Thus: \be P^{a_1a_2\ldots a_n}\propto
2^{V'-E'}\label{E'V'}\ee the same result obtained in
Sec.~\ref{app:Pratio} by another method.

\subsection{Periodic Boundary Conditions}
Now consider a honeycomb lattice of S=3/2's (no S=1/2's now) with
periodic boundary conditions.  This can be done in such a way that
every spin has 3 nearest neighbors and we take the AKLT ground
state.  Similar to the D=1 case, there can now be additional sets of
loops because we can form loops that wrap around the torus but don't
correspond to faces; see Fig.~\ref{fig:loopsPB}.  If a domain wraps
around the torus one way, but not the other, then the total number
of sets of loops, corresponding to the domain is: \be
N_{Li}=2^{F_i+1}.\ee To see this choose an arbitrary ``topological
loop'' within the domain going around the torus which doesn't
encircle any faces. There are now 2 sets of loops corresponding to
an arbitrary subset of faces, not using this topological loop and
the arbitrary subsets of faces together with the topological loop.
The construction of the set of loops corresponding to the
topological loop plus subset of faces is constructed by analogy with
the above construction. In cases where none of the faces share edges
with the topological loop the set of loops corresponds to the
topological loop plus the loops around the subset of
faces. In cases where one or more faces shares an edge with the
topological loop, the topological loop is modified to enclose each
such face; see Fig.~\ref{fig:loopsPB}b.

Finally, it is possible to have 2 topological loops in a domain,
going around the torus the two inequivalent ways; see
Fig.~\ref{fig:loopsPB}c. (In this case all other domains must be
topologically trivial.) We can grow the domain initially by drawing
these 2 topological loops without any faces. At this stage there are
3 closed loops, going around the torus one way or the other or using
all edges in the domain to go around both ways. Thus the number of
sets of loops is 4 at this stage.  After growing the entire domain
we how have: \be N_{Li}=2^{F_i+2}\ee since each set of loops
corresponds to a subset of faces, possibly combined with one of
these three topological loops. In general, we may associate a
winding number with each domain $W_i=0$, $1$ or $2$ with: \be
N_{Li}=2^{F_i+W_i}.\ee

It can be seen that the number of vertices and edges in each domain
obeys, in general: \be F_i+W_i=E_i-V_i+1.\ee This follows by
induction, as we grow the domain. At the step when we complete the
first topological loop we increase $E_i$ by 2 but $V_i$ by 1 and
$F_i$ by 0. Adding further faces respects \be \Delta F_i=\Delta
E_i-\Delta V_i\ee as before.  On the other hand, if we further grow
a second topological loop so that  the torus is encircled  both
directions, at the step where it goes around the torus in the second
direction we again increase $E_i$ by 2 but $V_i$ by 1 and $F_i$ by
0. Thus Eq. (\ref{E'V'}) remains true also with periodic boundary
conditions.


\begin{thebibliography}{99}
\bibitem{Betti} The Betti number of a graph $G$ is defined as $B\equiv
|E(G)|-|V(G)|+1$, and it counts the number of independent cylcles.
\end{thebibliography}
\end{document}